\documentclass[aps,prb,showpacs,twocolumn,superscriptaddress]{revtex4}
\usepackage{graphicx}
\usepackage{bm}
\usepackage{amsmath}

\def\be{\begin{equation}}       \def\ee{\end{equation}}
\def\bea{\begin{eqnarray}}      \def\eea{\end{eqnarray}}

\begin{document}

\title{SU(2) Non-Abelian Holonomy and Dissipationless Spin Current in Semiconductors}
\author{Shuichi Murakami}
\email[Electronic address: ]{murakami@appi.t.u-tokyo.ac.jp}
\affiliation{Department of Applied Physics, University of Tokyo,
Hongo, Bunkyo-ku, Tokyo 113-8656, Japan}
\author{Naoto Nagaosa}
\affiliation{Department of Applied Physics, University of Tokyo,
Hongo, Bunkyo-ku, Tokyo 113-8656, Japan} \affiliation{CERC, AIST
Tsukuba Central 4, Tsukuba 305-8562, Japan} \affiliation{CREST,
Japan Science and Technology Corporation (JST)}
\author{Shou-Cheng Zhang}
\affiliation{Department of Physics, McCullough Building, Stanford
University, Stanford CA 94305-4045}

\begin{abstract}
Following our previous work [S.\ Murakami, N.\ Nagaosa, S.\ C.\ Zhang, 
Science \textbf{301}, 1348 (2003)] on the
dissipationless quantum spin current, we present an exact quantum
mechanical calculation of this novel effect based on the linear
response theory and the Kubo formula. We show that it is possible
to define an exactly conserved spin current, even in the presence
of the spin-orbit coupling in the Luttinger Hamiltonian of p-type
semiconductors. The light- and the heavy-hole bands form two
Kramers doublets, and an $SU(2)$ non-abelian gauge field acts
naturally on each of the doublets. This quantum holonomy gives
rise to a monopole structure in momentum space, whose curvature
tensor directly leads to the novel dissipationless spin Hall
effect, i.e., a transverse spin current is generated by an
electric field. The result obtained in the current work gives a
quantum correction to the spin current obtained in the previous
semiclassical approximation.
\end{abstract}

\maketitle

\section{Introduction}
Spintronics, the science and technology of manipulating the spin
of the electron for building integrated information processing and
storage devices, showed great promise \cite{wolf2001}. Spintronics
devices also promises to access the intrinsic quantum regime of
transport, paving the path towards quantum computing. However,
many challenges remain in this exciting quest. Among them, purely
electric and dissipationless manipulation of the electron spin and
its quantum transport is one of the most important goals of
quantum spintronics.

In our previous work \cite{murakami2003}, we discovered a basic
law of spintronics, which relates the spin current and the
electric field by the response equation
\begin{equation}
j_j^i = \sigma_s \epsilon_{ijk} E_k \label{spin_response}
\end{equation}
where $j_j^i$ is the current of the $i$-th component of the spin
along the direction $j$ and $\epsilon_{ijk}$ is the totally
antisymmetric tensor in three dimensions. 
Sinova et al.\ \cite{sinova2003} found a similar effect in the two-dimensional 
n-type semiconductors with Rashba coupling.
This law is similar to
Ohm's law in electronics, and the spin conductivity $\sigma_s$ has
the dimension of the electric charge $e$ divided by the scale of
length. However, unlike the Ohm's law, this fundamental response
equation describes a purely topological and dissipationless spin
current. It is important to note here that the spin current is
even under the time-reversal operation ${\cal T}$. When the
direction of the arrow of time is reversed, both the direction of
the current and the spin are reversed and the spin current remains
unchanged. Since both the spin current and the electric field in
Eq.~(\ref{spin_response}) are even under time reversal ${\cal T}$,
the transport coefficient $\sigma_s$ is called ``reactive'' and
can be purely non-dissipative. This is in sharp contrast to the
Ohm's law
\begin{equation}
j_i=\sigma E_i \label{Ohm}
\end{equation}
relating the charge current $j_i$ to the electric field. In this
case, the charge current $j_i$ changes sign under time-reversal
${\cal T}$, while the electric field is even under ${\cal T}$.
Since the Ohm's law relates quantities of different symmetries
under time reversal ${\cal T}$, the charge conductivity $\sigma$
breaks the time-reveral symmetry and describes the inevitable
joule heating and dissipation. Quantum Hall current, which is
transverse to the electric field and dissipationless, has the
feature similar to Eq.~(\ref{spin_response}), but the
time-reversal symmetry is compensated by the external magnetic
field. Dissipationless current without time-reversal symmetry
breaking is extremely important and fundamental in solid-state
physics, the most celebrated example of which is the
superconducting current. It is described by the London equation,
\begin{equation}
j_i=\frac{\rho_s e^2}{mc} A_i \ \ , \ \
E_i=-\frac{1}{c}\frac{\partial A_i}{\partial t} \label{London}
\end{equation}
where the current $j_i$ is related to the vector potential $A_i$
instead of the electric field. In the London equation, both $j_i$
and $A_i$ are odd under time reversal ${\cal T}$, therefore, the
transport coefficient $\rho_s$, also called the superfluid
density, describes the reversible and dissipationless flow of the
supercurrent.

In summary, the dissipationless spin current discovered in
Ref.~\onlinecite{murakami2003} shares some basic features with the
superconducting current and the quantum Hall edge current, in the
sense that, 1) the spin Hall conductivity $\sigma_s$ is a
dissipationless or reactive transport coefficient, even under the
time-reversal operation ${\cal T}$; 2) the spin Hall conductivity
$\sigma_s$ can be expressed as an integral over all states below
the fermi energy, not only over states in the vicinity of the
fermi energy as in most dissipative transport coefficients.
Furthermore, just like the case of the quantum Hall
effect\cite{thouless1982,sundaram1999}, the contribution of each state to the
spin Hall conductivity $\sigma_s$ can be expressed entirely in
terms of the curvature of a gauge field in momentum space, which
in our case is non-abelian. The dissipationless spin current is
induced by the electric field through the spin-orbit coupling,
whose characteristic energy scale exceeds the room temperature in
many semiconducting materials.

Electronic structure of semiconductors with diamond structure
(e.g.\ Si, Ge) and zincblende structure (e.g.\ GaAs, InSb) are well
understood in terms of the ${\bf k} \cdot {\bf p}$ perturbation
theory. The top of the valence bands are at ${\bf k} ={0}$, i.e.,
$\Gamma$-point. They consist of the three p-orbitals $p_x,p_y,p_z$
with spin up and down. In the presence of the relativistic
spin-orbit coupling, these 6 states are split into four-fold
degenerate $S=3/2$ states and two-fold degenerate $S=1/2$ state.
Here $S$ denotes the total angular momentum of the atomic orbital,
obtained through the coupling of the orbital angular momentum $L$
and the spin angular momentum $s$. The second order perturbation
in the ${\bf k} \cdot {\bf p}$ results in the effective
Hamiltonian near ${\bf k} = {0}$, which is called Luttinger
Hamiltonian\cite{luttinger1956}:
\begin{eqnarray}
&&H_0=\sum_\mathbf{k} c_{\mu,\mathbf{k}}^{\dagger}
H_{\mu\nu}(\mathbf{k})c_{\nu,\mathbf{k}}\nonumber \\
&&H_{\mu\nu}(\mathbf{k})=\frac{1}{2m}
\left((\gamma_{1}+\frac{5}{2}\gamma_{2})k^{2}-
2\gamma_{2}(\mathbf{k}\cdot \mathbf{S}
)^{2}\right)_{\mu\nu} \label{Luttinger}
\end{eqnarray}
where $\mathbf{k}=(k_x,\ k_y,\ k_z)$, $\mathbf{S}=(S^x,\ S^y,\
S^z)$, and $k=|\mathbf{k}|$. The explicit form of the matrices
$S^i$ $(i=1,2,3)$ is given in Appendix \ref{appendix-matrix}. For
simplicity, we have put $\gamma_{2}=\gamma_{3}$ in the original
Luttinger Hamiltonian; most of subsequent discussions are also
applicable to more general cases with $\gamma_{2}\neq\gamma_{3}$.

\begin{figure}[h]
\includegraphics[scale=0.8]{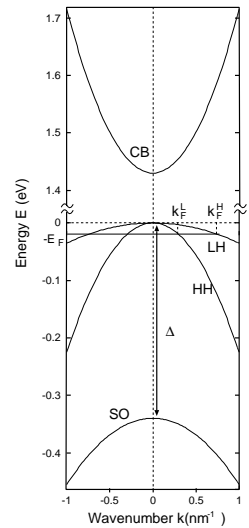}
\caption{
Schematic picture of the band structure of GaAs near $\mathbf{k}=0$.
CB means the conduction band, HH the heavy-hole band, LH the 
light-hole band, and
SO the split-off band, respectively. When the small inversion symmetry 
breaking is neglected, all of them are doubly degenerate.
The Fermi energy shown in the figure corresponds to the hole 
density 
$\sim 10^{19}\text{cm}^{-3}$.
The splitting $\Delta=0.34$eV at $\mathbf{k}=0$ 
between the LH,HH and SO bands are due to the spin-orbit interaction.
The HH and LH bands are degenerate at $\mathbf{k}=0$, giving rise to 
a monopole in the gauge field as discussed in Section \ref{kubo_calculation}} 
\label{fig-1}
\end{figure}
On the other hand, the conduction bands are made out of the
s-orbital and hence doubly degenerate. When we neglect a small
effect due to broken inversion symmetry, this degeneracy is not
lifted due to the Kramers theorem. Therefore the effect of the
spin-orbit interaction is small in the conduction band, although
the Rashba effect \cite{rashba1960,bychkov1984,
nitta1997,datta1990} is induced by the electric field near e.g.
the interface structure. The spin Hall current in the Rashba
system has recently been discussed by Sinova et
al.~\cite{sinova2003}. These authors showed that dissipationless
and intrinsic spin Hall current can take an universal value in
this system. The position of the conduction band minima depends on
the material. For example, they are located at general points
along the axis between the $\Gamma$- and the $X$-points in Si,
while they at at the $L$-points in Ge. We will focus on the
valence bands below because of the intrinsically strong spin-orbit
interaction.

Although the band structure of semiconductors with spin-orbit
coupling has been understood for many years, only recently has it
been recognized that the gauge field and its
curvature in the momentum space made out
of the Bloch wavefunction play important roles in the transport
properties of electrons in solids. The gauge field
is defined in terms of the
Bloch state $| n \mathbf{k} \rangle$ as
\begin{equation}
A_{n i }(\mathbf{k}) = -i \left\langle n \mathbf{k}
 \left|\frac{\partial}{ \partial k_i}
\right|n\mathbf{k}\right\rangle,
\end{equation}
where $n$ is a band index.
It represents the inner product of the two Bloch wavefunctions
infinitesimally separated in $\mathbf{k}$-space.
This gauge field $A_{ni}(\mathbf{k})$ describes
topological structure of Bloch wavefunctions in the momentum space
\cite{thouless1982,sundaram1999},
and plays an important role in transport properties
\cite{matl1998,ye1999, chun2000,ohgushi2000,
taguchi2001,jungwirth2002,taguchi2003,fang2003,yasui2003}
and
in magnetic superconductors \cite{murakami2003b}.
In particular, this gauge field
is related to the transverse conductivity $\sigma_{xy}$ as
\begin{equation}
\sigma_{xy} = -\frac{e^{2}}{h}\sum_{n, \mathbf{k}} n_{F}(
\epsilon_{n}(\mathbf{k})) B_{nz}(\mathbf{k}),
\end{equation}
where $B_{ n z}(\mathbf{k}) =F_{n,xy}(\mathbf{k})=
 \frac{\partial A_{n y}
(\mathbf{k})}{\partial k_{x}} -\frac{\partial A_{n
x}(\mathbf{k})}{\partial k_{y}}$ is the $z$-component of the 
field strength made from $A_{n i}(\mathbf{k})$, and
$n_{F}(\epsilon_{n}(\mathbf{k}))$ is the Fermi distribution of fr the
$n$-th band with energy $\epsilon_{n}( \mathbf{k})$. This formula
\cite{thouless1982,sundaram1999} is the foundation of the integer quantum Hall
effect (QHE), and further applied to the anomalous Hall effect
(AHE) in ferromagnetic metals \cite{matl1998,ye1999,
chun2000,ohgushi2000, taguchi2001,jungwirth2002,taguchi2003,
fang2003,yasui2003}. Especially in the magnetic semiconductors
(Ga,Mn)As \cite{ohno1998}, the calculation \cite{jungwirth2002} by
the formula Eq.~(\ref{London}) well explains the experimental
results quantitatively, giving some credit that the AHE is mostly
of intrinsic origin rather than extrinsic origins, e.g., skew
scattering and/or side-jump mechanism.  However in the presence of
the time-reversal symmetry, the d.c. transverse conductivity
$\sigma_{xy}$ vanishes, and the topological structure of the Bloch
wavefunctions has not been systematically studied in the context
of transport theory. As we will show below, an even more beautiful
and nontrivial quantum topological structure is hidden in the
valence-band structure in the paramagnetic state, which is
analogous to the fermionic quasi-particles in the SO(5)
theory\cite{demler1999}. This is also motivated by the recent work
by one of the present authors on the generalization of the quantum
Hall effect into four dimensions in terms of the SO(5)
symmetry\cite{zhang2001}. In this paper, we shall show that the
SO(5) group structure of the $S=3/2$ Bloch states provides a
natural description of the non-abelian $SU(2)$ holonomy and its
curvature in momentum space. This gauge structure underlies the
dissipationless, topological spin current in hole-doped
semiconductors.

In the presence of the spin-orbit interaction, the conventionally
defined total spin operator is not conserved, and it is nontrivial
to define the spin current in this case. Our formalism resolves
this issue by discovering conserved quantities in the Luttinger
Hamiltonian (\ref{Luttinger}) and by defining associated conserved
spin currents. These quantities have clear physical and geometric
meanings. The exact quantum calculation of the conserved spin Hall
conductivity $\sigma_s$ is performed in terms of the Kubo formula,
and the results can be expressed entirely in terms of the
non-abelian gauge curvature in momentum space. 
Our fully quantum mechanical results identify the 
quantum correction to the previous semiclassical result\cite{murakami2003}
in terms of the wave-packet formalism.

The plan of this paper is as follows. In section \ref{definition},
we reformulate the Luttinger Hamiltonian in terms of the SO(5)
algebra and give the definition of the conserved spin current.
Based on this definition, the calculation of the spin Hall
conductivity in terms of Kubo formula is presented in section
\ref{kubo_calculation}, where the geometrical meaning is stressed,
and the comparison with the previous semiclassical result is
given. Section \ref{conclusion} is devoted to conclusions.
Throughout the paper, we take the unit $\hbar=1$, $e=1$.

\section{Definition of the spin current}
\label{definition} 
\begin{figure}[h]
\includegraphics[scale=0.45]{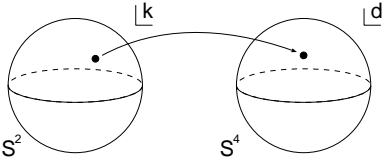}
\caption{Mapping from the three-dimensional $\mathbf{k}$ vector
space to the five-dimensional $\mathbf{d}$ vector space. The gauge
structures of the spin current are associated with the magnetic
monopoles located at the origins of the $\mathbf{k}$ and
$\mathbf{d}$ spaces. } \label{fig-2}
\end{figure}

The Luttinger Hamiltonian (\ref{Luttinger}) has
two eigenvalues:
\begin{equation}
E_{L}(k)=\frac{\gamma_{1}+2\gamma_{2}}{2m}k^{2} \ \ , \ \
E_{H}(k)=\frac{\gamma_{1}-2\gamma_{2}}{2m}k^{2} \label{E-values}
\end{equation}
corresponding to the light-hole (LH) and the heavy-hole (HH)
bands. Each eigenvalues are doubly degenerate, due to the Kramers
theorem based on the time-reversal symmetry. For a fixed value of
$\mathbf{k}$, let $P^{L}(\mathbf{k})$ denote a projection onto the
two-dimensional subspace of states of the LH band. We also define
$P^{H}(\mathbf{k})$ similarly. These operators are written as
\begin{equation}
P^{L}=\frac{9}{8}-\frac{1}{2k^{2}}(\mathbf{k}\cdot\mathbf{S})^{2}
\ \ ,\ \ P^{H}=1-P^{L}\label{k-projector}
\end{equation}
They obviously satisfy
\begin{equation}
P^{L}P^{H}=0=P^{H}P^{L},\ \ (P^{L})^{2}=P^{L},\ \
(P^{H})^{2}=P^{H}
\label{k-projector-condition}
\end{equation}
In terms of these projectors, the Luttinger Hamiltonian can be
expressed as
\begin{equation}
H=\sum_{\mathbf{k}}(E_{H}(\mathbf{k})P^{H}(\mathbf{k})+E_{L}
(\mathbf{k})P^{L}(\mathbf{k})). \label{Hamiltonian-projection}
\end{equation}
{}From this projector form of the Hamiltonian, we see that the LH
and the HH bands are each two-fold degenerate, and there is an
SU(2) rotation symmetry acting on each band. Combining the LH and
the HH bands, there is an SO(4)$=$SU(2)$\times$SU(2) symmetry at
every $\mathbf{k}$ point. In this section, we shall develop the
mathematical framework in which this SO(4) symmetry is made
manifest, and this symmetry is used to define the conserved spin
current. Since $P^{H}$ and $P^{L}$ depend on $\mathbf{k}$, the
quantization axis for each SU(2) varies as a function of
$\mathbf{k}$. When $\mathbf{k}$ is adiabatically changed along a
closed circuit, the fermionic wave function in general does not
return to itself; in fact, the final wave function is related to
the starting wave function by an SU(2) transformation within each
band. Therefore, this problem is a natural generalization of
Berry's U(1) phase\cite{berry1984} to the case of SU(2)
holonomy\cite{wilczek1984,avron1988,zee1988,mathur1991,shankar1994,arovas1998,demler1999}.
In particular, Demler and Zhang\cite{demler1999} developed a
formalism of the SU(2) non-abelian holonomy in terms of the SO(5)
Clifford algebra, which we shall adopt throughout this paper. Upon
expanding the $(k_i S^i)^{2}$ term in the Hamiltonian
(\ref{Luttinger}), we obtain a product of two quadratic forms, one
of the form $\xi_a^{ij} k_i k_j$ and another of the form
$\xi_a^{ij} S^i S^j$, where $\xi_a^{ij}$ is a symmetric matrix. A
3$\times$3 symmetric matrix can be further decomposed into one
trace and five traceless parts. The trace part has the same form
as the first term in the Hamiltonian (\ref{Luttinger}), and
cancels the $\gamma_2$ contribution by construction. The remaining
five traceless symmetric ($\xi_a^{ij}=\xi_a^{ji}, \xi_a^{ii}=0$)
combination of $\xi_a^{ij} S^i S^j$ can be identified with the
Clifford algebra of the Dirac $\Gamma$ matrices, with the
identification
\begin{equation}
\Gamma^a=\xi_a^{ij}\{S^i, S^j\}\ ,\ \ \{\Gamma^a,
\Gamma^b\}=2\delta_{ab} \label{gamma}
\end{equation}
The explicit forms of $S^i, \Gamma^a$ and $\xi_a^{ij}$ are given
in Appendix \ref{appendix-matrix}.
In terms of the $\Gamma$
matrices, the Luttinger Hamiltonian (\ref{Luttinger}) takes the
elegant form
\begin{equation}
H(k)=\epsilon(\mathbf{k})+\frac{\gamma_{2}}{m}d_{a}\Gamma^{a},
\label{Hamiltonian}
\end{equation}
where
\begin{eqnarray}
&&\epsilon(\mathbf{k})=\frac{\gamma_{1}}{2m}k^{2},\ d_a(\mathbf{k}
)=-3\xi_a^{ij} k_i k_j,\nonumber \\
&& d_{1}=-\sqrt{3}k_{y}k_{z},\ d_{2}=-\sqrt{3} k_{x}k_{z},\
d_{3}=-\sqrt{3} k_{x}k_{y},\nonumber \\
&& d_{4}=-\frac{\sqrt{3}}{2}(k_{x}^{2}-k_{y}^{2}),\nonumber \\
&& d_{5}=-\frac{1}{2}(2k_{z}^{2}-k_{x}^{2}-k_{y}^{2}).
\label{d}\end{eqnarray} We recognize that the vector components of
$d_a(\mathbf{k})$ are nothing but the five d-wave combinations in
the $\mathbf{k}$ space. The five-dimensional vector $\mathbf{d}$
provides a mapping from the three-dimensional $\mathbf{k}$ space
to the five-dimensional $\mathbf{d}$ space (Fig.~\ref{fig-2}).
Since the Luttinger Hamiltonian depends on $\mathbf{k}$ only
through $\mathbf{d(k)}$, we can perform all calculations in the 5D
$\mathbf{d}$ space, and finally project back onto the 3D
$\mathbf{k}$ space. This formalism enables a unified treatment
for the anisotropic Luttinger Hamiltonian, and more importantly,
reveals the deep connection to the four-dimensional quantum 
Hall effect (4DQHE) \cite{zhang2001}. Here and
henceforth we adopt the convention that indices appearing twice
are summed over.

The eigenvalues of (\ref{Hamiltonian}) are
$E_{L}(\mathbf{k})=\epsilon(\mathbf{k})+\frac{\gamma_{2}}{m
}d(\mathbf{k})$ and
$E_{H}(\mathbf{k})=\epsilon(\mathbf{k})-\frac{\gamma_{2}}{m}d(\mathbf{k})$,
where $d(\mathbf{k})=\sqrt{d_{a}d_{a}}=k^{2}$. They are of course
the same as (\ref{E-values}). In terms of the $\Gamma$ matrices,
the projection operators (\ref{k-projector}) can be expressed as
\begin{equation}
P^{L}=\frac{1}{2}(1+\hat{d}_{a}\Gamma^{a}) \ \ , \ \,
P^{H}=\frac{1}{2}(1-\hat{d}_{a}\Gamma^{a}),\label{d-projector}
\end{equation} where $\hat{d}_{a}=
d_{a}/d$. The $\Gamma$ matrices are convenient for subsequent
calculations, since a product of any number of $\Gamma$ matrices
can be easily reduced to a linear combination of $1$, $\Gamma^{a}$
and $\Gamma^{ab}=\frac{1}{2i} [\Gamma^{a},\Gamma^{b}]$. The five
$\Gamma^a$ matrices contain the most general quadratic terms of
the spin operator $S^i$, while the ten $\Gamma^{ab}$ matrices
contain both the three spin operators $S^i$ and the seven cubic,
symmetric and traceless combinations of spin operators of the form
$S^i S^j S^k$, as discussed in Appendix \ref{appendix-matrix}.
These ten spin
operators are generated under the Heisenberg time evolutions of
the single spin operators, and it is natural to group them all
into a unified object. For $\gamma_2=0$, $\Gamma^{ab}$ commutes
with the Hamiltonian and generates an SO(5) symmetry group of the
Hamiltonian. (In fact, the Hamiltonian has a higher, SU(4)
symmetry in this case). For a given $\mathbf{k}$, a fixed $\mathbf{d}$ vector
singles out a particular direction in the five-dimensional $\mathbf{d}$
vector space, and the second term in (\ref{Hamiltonian}) breaks
the SO(5) symmetry to an SO(4) symmetry. This is nothing but
SO(4)$=$ SU(2)$\times$SU(2) symmetry which we discussed earlier.
In this way, we see that the SO(5) formalism gives an elegant
geometric interpretation of the SU(2)$\times$SU(2) symmetry of the
LH and the HH bands. It is in fact a subgroup of SO(5) rotation
which keeps a fixed $\mathbf{d}$ vector invariant. As we shall see later in
the paper, and in Appendix \ref{appendix-gauge}, the 3D monopole
structure in $\mathbf{k}$ space can be best understood in terms of the
monopole structure in the 5D $\mathbf{d}$ space.

To find the conserved quantities, let us define the conserved spin
density explicitly as
\begin{equation}
\rho^{ab}(\mathbf{p},\mathbf{q})=
c_{\mathbf{p}+\frac{\mathbf{q}}{2},\mu}^{\dagger}
P_{ab,cd}(\mathbf{p},\mathbf{q})\Gamma_{\mu\nu}^{cd}
c_{\mathbf{p}-\frac{\mathbf{q}}{2},\nu}, \label{rho}
\end{equation}
where $P_{ab,cd}=P_{cd,ab}=-P_{ba,cd}=-P_{ab,dc}$. For the
conservation of the spin, we require that $
-i\dot{\rho}^{ab}(\mathbf{p},\mathbf{q})=[H,\rho^{ab}(\mathbf{p},
\mathbf{q})]$ is proportional to $\mathbf{q}$ for small
$|\mathbf{q}|$. This is realized by imposing a condition on
$P_{ab,cd}$ as
\begin{equation}
P_{ab,cd}(\mathbf{p},\mathbf{q})\left(
d_{c}\left(\mathbf{p}+\frac{\mathbf{q}}{2}\right)+d_{c}\left(
\mathbf{p}-\frac{\mathbf{q}}{2}\right)\right)
=0
\end{equation}
or equivalently,
\begin{equation}
P_{ab,cd}(\mathbf{p},\mathbf{q})\left(
d_{a}\left(\mathbf{p}+\frac{\mathbf{q}}{2}\right)
+d_{a}\left(\mathbf{p}-\frac{\mathbf{q}}{2}\right)\right) =0.
\label{xiabcd-condition}\end{equation} From these relations it
follows that
\begin{equation}
\rho^{ab}(\mathbf{p},\mathbf{q})\left(
d_{a}\left(\mathbf{p}+\frac{\mathbf{q}}{2}\right)
+d_{a}\left(\mathbf{p}-\frac{\mathbf{q}}{2}\right)\right)
=0.
\end{equation}
There are five such linear equations, but only four of them are
linearly independent because of the antisymmetry of $\rho^{ab}$.
Originally, $\rho^{ab}$ has 10 degrees of freedom, subtracting 4
constraints gives the remaining 6 degrees of freedom, exactly the
same as the number of generators in SO(4) algebra. Therefore, the
projection operator $P_{ab,cd}$ projects the full SO(5) symmetry
generators into the SO(4) subspace which is orthogonal to a given
direction of $\mathbf{d}$.

In the limit $\mathbf{q}=0$, (\ref{xiabcd-condition}) is satisfied
by
\begin{eqnarray}
&& P_{ab,cd}
=\frac{1}{2}\left(\delta_{ac}\delta_{bd}-\delta_{ad}\delta_{bc}
+\delta_{ad}\hat{d}_{b}\hat{d}_{c}\right.\nonumber \\
&&\ \ \ -\left.\delta_{bd}\hat{d}_{a}\hat{d}_{c}
-\delta_{ac}\hat{d}_{b}\hat{d}_{d}
+\delta_{bc}\hat{d}_{a}\hat{d}_{d}\right). \label{P-projector}
\end{eqnarray}
It satisfies $ P_{ab,cd}
P_{cd,ef}= P_{ab,ef}$, implying that  
it is properly normalized as a projection.
It is interesting to note that $P_{ab,cd}$
can also be expressed as
\begin{equation}
P_{ab,cd}=f_{aba'b'}f_{a'b'cd}, \
f_{abcd}=\frac{1}{2}\epsilon_{abcde}\hat{d}_{e}.
\label{Pabcd}
\end{equation}
Inserting
(\ref{P-projector}) into the spin density (\ref{rho}), we obtain
\begin{eqnarray}
&&\rho^{ab}= \sum_{\mathbf{k}}c^{\dagger}_{\mathbf{k}\mu}
P_{ab,cd}(\mathbf{k})
(\Gamma^{cd})_{\mu\nu}c_{\mathbf{k}\nu}\nonumber \\
&&\ \ \ =\sum_{\mathbf{k}}c^{\dagger}_{\mathbf{k}\mu}
\left(\Gamma^{ab}-\frac{i}{2}[\hat{d}_{a}\Gamma^{b}
-\hat{d}_{b}\Gamma^{a},
\hat{d}_{f}\Gamma^{f}]\right)_{\mu\nu} c_{\mathbf{k}\nu}.\ \ \ 
\end{eqnarray}
Because $ [\Gamma^{ab}, d_{e}\Gamma^{e}]=
2i(d_{a}\Gamma^{b}-d_{b}\Gamma^{a})$, we get
\begin{eqnarray}
&&\rho^{ab}=\sum_{\mathbf{k}}c^{\dagger}_{\mathbf{k}\mu}
\left(\Gamma^{ab}-\frac{1}{4} [[\Gamma^{ab},
\hat{d}_{e}\Gamma^{e}],\hat{d}_{f}\Gamma^{f}]
\right)_{\mu\nu}
c_{\mathbf{k}\nu}\nonumber \\
&&\ \ =\sum_{\mathbf{k}}c^{\dagger}_{\mathbf{k}\mu}
\left(P^{L}\Gamma^{ab}P^{L}+ P^{H}\Gamma^{ab}P^{H}
\right)_{\mu\nu} c_{\mathbf{k}\nu}. \label{PLPH}
\end{eqnarray}
Thus it corresponds to projecting out the inter-band matrix
elements of $\Gamma^{ab}$. The conservation of $\rho^{ab}$ becomes
manifest in (\ref{PLPH}), because the Hamiltonian is diagonal in
each subspace, i.e. the LH or the HH band.

The equation of continuity determines the uniform spin current to
be
\begin{eqnarray}
&&J_{i}^{ab}=
\frac{1}{2}\sum_{\mathbf{k},\mu,\nu}c_{\mathbf{k}\mu}^{\dagger}\left\{
\frac{\partial H}{\partial k_{i}},\  P_{ab,cd}\Gamma^{cd}
\right\}_{\mu\nu}c_{\mathbf{k}\nu}
\nonumber \\
&&\ \ 
=\frac{1}{2}\sum_{\mathbf{k},\mu,\nu}c_{\mathbf{k}\mu}^{\dagger}
\left\{ \frac{\partial \epsilon}{\partial k_{i}} +
\frac{\gamma_{2}}{m}\frac{\partial d_{f}}{\partial
k_{i}}\Gamma^{f} , \right.\nonumber \\
&&\makebox[2cm]{}
\left. P^{L}\Gamma^{ab}P^{L}+ P^{H}\Gamma^{ab}P^{H}
 \right\}_{\mu\nu} c_{\mathbf{k}\nu}.
\label{spin-current}\end{eqnarray}

To connect this spin current with the physical spin current in
Ref.~\onlinecite{murakami2003}, we define a tensor $\eta_{ab}^{i}$
by $S^i=\eta_{ab}^{i}\Gamma^{ab}=\frac{1}{2i}
\eta_{ab}^{i}[\Gamma^{a},\Gamma^{b}].$  Explicit forms 
and properties of $\eta_{ab}^{i}$ are summarized in Appendix 
\ref{appendix-matrix}.
By contracting with
$\frac{1}{3}\eta^{k}_{ab}$, the conserved spin takes the form
\begin{eqnarray}
&&S^{l}_{\text{(c)}}=\frac{1}{3}\eta^{l}_{ab}\rho^{ab}\nonumber \\
&& \ \ \
=\frac{1}{3}\sum_{\mathbf{k}}c_{\mathbf{k}\mu}^{\dagger}
\left(P^{L}S^{l}P^{L}+ P^{H}S^{l}P^{H} \right)_{\mu\nu}
c_{\mathbf{k}\nu}. \label{sk-c}
\end{eqnarray}
The subscript $_{(c)}$ denotes the fact that this spin current is
conserved.
Here, we inserted a factor of $\frac{1}{3}$, because
in the LH and HH bands (i.e. $S=3/2$ subspace), the expectation value of
the spin angular momentum is one-third of that of the total
angular momentum $S^{k}$.
Thus, Eq.~(\ref{sk-c})
corresponds to neglecting interband matrix elements of the spin
angular momentum.
In a matrix form, the corresponding
conserved spin current is
\begin{equation}
J^{l}_{i\text{(c)}}= \frac{1}{3}\cdot\frac{1}{2}\left\{
\frac{\partial H}{\partial k_{i}},\ P^{L}S^{l}P^{L}
+P^{H}S^{l}P^{H} \right\}. \label{anticommutator}
\end{equation}

\section{Kubo formula calculation of the spin current}
\label{kubo_calculation}
\subsection{Difficulties with the conventional definition of the
non-conserved spin current} In order to calculate the spin current
response based on Kubo formula, we should first define the ``spin
current operator''. The conventional definition of the spin
current, with spin along the $l$ axis flowing along the $i$ axis,
is given by
\begin{equation}
J_{i}^{l}=\frac{1}{3}\cdot \frac{1}{2}\left\{\frac{\partial
H}{\partial k_i}, \ S^{l}\right\}. \label{conventional}
\end{equation}
Because the spin $S^{l}$ is not conserved,
$J_{i}^{l}$ does not satisfy the equation of
continuity without any source term. Before
presenting the full calculation based on the conserved spin
current discussed in the previous section, we first calculate the
linear response of this non-conserved spin current to the applied
electric field and then comment on its difficulties. The Kubo
formula gives
\begin{eqnarray}
&&Q_{ij}^{l}(i\nu_{m}) =-\frac{1}{V}\int_{0}^{\beta} \langle \hat{T}
J_{i}^{l}(u){J}_{j}\rangle e^{i\nu_{m}u}du
\nonumber \\
&&\ \ \ =\frac{1}{V\beta}\sum_{k,n} \text{tr}\left(
J_{i}^{l}G(\mathbf{k},i(\omega_{n}+\nu_{m}))J_{j} G
(\mathbf{k},i\omega_{n})\right),\ \ 
\label{Qijkinul}\end{eqnarray}
where $\nu_{m}=2\pi m/\beta$ ($m$: integer), $\omega_{n}
=(2n+1)\pi /\beta$ ($n$: integer), $\beta=1/k_{B}T$,
$\hat{T}$ in (\ref{Qijkinul}) represents the time-ordering,
\begin{equation}
J_{j}=
\sum_{\mathbf{k},\mu,\nu}c_{\mathbf{k}\mu}^{\dagger}\left(
\frac{\partial \epsilon}{\partial k_{j}} +\sum_{h} \frac{\partial
d_{h}}{\partial k_{j}} \Gamma^{h}
\right)_{\mu\nu}c_{\mathbf{k}\nu},
\end{equation}
and $G(\mathbf{k},i\omega_{n})$ is the Matsubara Green's function,
given in (\ref{Green}).

In the clean case, the summation over $\omega_{n}$ can be
calculated by a contour integral. In the trace operation in the
above equation, only the terms of products of four or five
$\Gamma$ matrices are nonzero, and the result is
\begin{eqnarray}
&&Q_{ij}^{l}(\omega+i\delta)= \frac{4i\omega}{V}\sum_{\mathbf{k}}
\frac{n_{F}(\epsilon_{L})-n_{F}(\epsilon_{H})}{d(\omega^{2}-4\gamma_{2}^{2}
d^{2}/m^{2})}\nonumber
\\
&&\ \ \cdot \left(\frac{\gamma_{2}}{m}\right)^{2}
\left(\frac{\gamma_{1}}{2\gamma_{2}}\epsilon_{ljk}k^{2}k_{i}k_{k}
-\epsilon_{ijk}k^{2}k_{k}k_{l}\right). \label{Qijk}
\end{eqnarray}
where $\epsilon_{L}(\mathbf{k})
=E_{L}(\mathbf{k})-\mu$ and
$\epsilon_{H}(\mathbf{k})=E_{H}(\mathbf{k})
-\mu$ are one-particle energies for the two bands,
measured form the chemical potential $\mu$.

Therefore, in the static limit the linear response is given by
\begin{eqnarray}
&&\sigma_{ij}^{l}=\lim_{\omega\rightarrow
0}\frac{Q_{ij}^{l}(\omega)}{-i\omega}
\nonumber \\
&&\ \ = \frac{1}{3V}\sum_{\mathbf{k}}
\frac{n_{F}(\epsilon_{L})-n_{F}(\epsilon_{H})}{k^{2}}
\epsilon_{lji}\left(\frac{\gamma_{1}}{2\gamma_{2}}+1 \right)
\nonumber
\\
&&\ \ = \frac{1}{6\pi^{2}}\epsilon_{ijl} (k_{F}^{H}-k_{F}^{L})
\left(\frac{\gamma_{1}}{2\gamma_{2}}+1 \right). 
\label{kubo-nonconserved}
\end{eqnarray}
This result  $\sigma_{ij}^{l}$ in Eq.~(\ref{kubo-nonconserved})
does not vanish in the limit of $\gamma_2 \to 0$, i.e., the
absence of the spin-orbit coupling. 
%
It is not a contradiction, 
because the two limits $\gamma_2
\to 0$ and $\omega \to 0$ cannot be exchanged in Eq.~(\ref{Qijk}),
and Eq.~(\ref{kubo-nonconserved}) is the one which is valid in the d.c.
limit, when $\frac{\gamma_2}{m}k^{2} \gg \omega$.
We have learned that Hu, Bernevig and Wu have also 
obtained a similar result independently \cite{hu2003}.
We note that this result (\ref{kubo-nonconserved})
is reproduced by wave-packet dynamics in 
ref.\ \onlinecite{culcer2003}.

The conventional definition of the spin current (\ref{conventional})
is physically admissible, as is usually adopted. However, its mathematical
meaning as a ``current'' is ill-defined.
A ``current'' is always associated with a corresponding conserved
quantity. A ``current'' is then
defined by using the Noether's theorem, or equivalently, by the
equation of continuity. Since the conventionally defined spin current
is not conserved for the Luttinger Hamiltonian due to the spin-orbit
coupling, we shall use the the conserved spin current constructed in
the previous chapter.

There are also physical reasons to take this conserved spin current.
Generally speaking, there must be some reason for a quantity to
have slow dynamics and to contribute to the low frequency
response. One is a conservation law and the other is a
critical slowing down. In the present context, the latter is
irrelevant and we need to look for a conserved current as we
have done in the preceding chapter. When we separate the spin
into the conserved and the nonconserved parts,
the nonconserved part
\begin{equation}
S^{l}_{i\text{(n)}}= \frac{1}{3}
\left(P^{L}S^{l}P^{H}
+P^{H}S^{l}P^{L}\right)
\end{equation}
has an oscillating factor in time $e^{\pm i(E_{L}-E_{H})t}$
in the Heisenberg picture. Its frequency is $E_{L}-E_{H}$ 
and is nominally 0.1-1 eV or 1-10 fsec.
As we are observing spins averaged 
over the time-scale much longer than 1-10fsec,
the only remaining part is the conserved part.
Thus, in addition to mathematical soundness,
the conserved part of the spin current
automatically takes into account this averaging over time.
In the next section we shall 
calculate the d.c. response in terms of
the conserved part of the spin current.

\subsection{Kubo formula calculation for the conserved spin current}
In contrast with the previous approach, the approach using
conserved spin $S^{l}_{\text{(c)}}$ gives well-defined and
conserved spin current (\ref{anticommutator}). This approach is
equivalent to neglect interband matrix elements of spin operators
$S^{l}$, as seen from (\ref{sk-c}). This is justified in
calculation of spin current because of the following reason. Let
us consider the problem in a semiclassical way. Two wave-packets
in different bands are moving with different velocities, and they
will move apart inside the sample. Meanwhile, in the sample there
are sources causing decoherence between wave-packets, e.g.
inelastic scattering. This decoherence effect smears out the
interband matrix elements. Therefore, in the measurement of the
spin current, what is measured is only an intraband matrix element
of spin carried by a hole coming out of the sample. Thus in the
measurement of the spin current, we should consider only the
intraband matrix element of $S^{l}$. This is in contrast with
calculation of 
susceptibility, where intraband matrix elements of spin gives
significant contributions.

By applying the electric field, this (conserved) spin current is
induced by spin-orbit coupling. Let us calculate this linear
response according to Kubo formula. Hence we shall calculate
\begin{eqnarray}
&&Q_{ij}^{ab}(i\nu_{m})= -\frac{1}{V}\int_{0}^{\beta} \langle \hat{T}
J_{i}^{ab}(u){J}_{j}\rangle e^{i\nu_{m}u}du
\nonumber \\
&&\ \ =\frac{1}{V\beta}\sum_{k,n} \text{tr}\left(
J_{i}^{ab}G(\mathbf{k},i(\omega_{n}+\nu_{m}))J_{j} G
(\mathbf{k},i\omega_{n})\right). \ \ \ \ \ \label{conserved-Kubo}
\end{eqnarray}
By evaluating the summation over $\omega_{n}$ and taking the trace
as presented in Appendix \ref{appendix-kubo}, we get
\begin{equation}
Q_{ij}^{ab}(i\nu_{m})=\frac{-16\nu_{m}}{V}\left(
\frac{\gamma_{2}}{m}\right)^{2} \sum_{\mathbf{k}}
\frac{n_{F}(\epsilon_{L})-n_{F}(\epsilon_{H})}{(i\nu_m)^{2}-4\gamma_{2}^{2}
d^{2}/m^{2}} d^{2}G^{ab}_{ij},
\end{equation}
where \begin{equation} G^{ab}_{ij}=\frac{1}{4d^{3}}
\epsilon_{abcde}d_{c}\frac{\partial d_{d}}{\partial k_{i}
}\frac{\partial d_{e}}{\partial k_{j}} \label{Gij}\end{equation}
is a purely geometric tensor. In the static limit we have,
\begin{eqnarray}
&&\sigma_{ij}^{ab} = \lim_{\omega\rightarrow 0}
\frac{Q_{ij}^{ab}(\omega)}{-i\omega}\nonumber
\\
&&\ \ \ =\frac{4}{V}\sum_{\mathbf{k}}
(n_{L}(\mathbf{k})-n_{H}(\mathbf{k}))
G^{ab}_{ij}\nonumber \\
&&\ \ \ =\frac{4}{V}\sum_{\mathbf{k}}
(n_{L}(\mathbf{k})-n_{H}(\mathbf{k}))
(F_{ij}^{L,ab}-F_{ij}^{H,ab}), \label{sigmaijab}\end{eqnarray}
where $n_{L}=n_{F}(\epsilon_{L})$, $n_{H}=n_{F}(\epsilon_{H})$ are
the Fermi functions of the LH and the HH bands. Here
$F^{L,ab}_{ij}$ and $F^{H,ab}_{ij}$ are non-abelian gauge field
strengths, i.e. curvature of the 
gauge field in the LH and the HH bands, and their definition and
formulae are given in Appendix \ref{appendix-gauge}. In contrast
to the result of the non-conserved spin current, the conductivity
of the conserved spin current (\ref{sigmaijab}) is expressible in
terms of purely geometric quantities. Here we note that 
$G^{ab}_{ij}$ as given in
(\ref{Gij}) is similar to the $\theta$ term in the
(1$+$1)-dimensional O(3) nonlinear $\sigma$-model, which takes the
form of
\begin{equation}
\epsilon_{\alpha\beta}\epsilon_{ijk}n_{i} \frac{\partial
n_{j}}{\partial  k_{\alpha}} \frac{\partial n_{k}}{\partial
k_{\beta}}.
\end{equation}
In fact, (\ref{Gij}) describes the mapping of an area form from
the three-dimensional (${R}^{3}$) $\mathbf{k}$ space to the
five-dimensional ($R^{5}$) $d_{a}(\mathbf{k})$ space . An area
element on $R^{3}$ has 3 orientations $dk_i\wedge dk_{j}$, while
an area element on $R^{5}$ has 10 orientations, $d(d_{a})\wedge
d(d_{b})$. Our formula describes the Jacobian of the area map. Out
of the 10 possible orientations of an area form in $R^{5}$, the
$f_{abcd}=\frac{1}{2}\epsilon_{abcde}\hat{d}_{e}$ tensor in
(\ref{Gij}) selects 6 orientations which are locally transverse to
$\hat{d}_{a}$. Geometric properties of the $G^{ab}_{ij}$ tensor
are further summarized in Appedix \ref{appendix-gauge}.

By substituting the formula (\ref{Gabij}) for $G^{ab}_{ij}$,
we get
\begin{equation}
\sigma_{ij}^{ab}
=\frac{4}{5\pi^{2}}\eta_{ab}^{l}\epsilon_{ijl}(k_{F}^{H}-k_{F}^{L}).
\end{equation}
By contracting with $\frac{1}{3}\eta_{ab}^{l}$, the linear
response of the corresponding current is
\begin{equation}
\sigma_{ij\text{(c)}}^{l}
\equiv\frac{1}{3}\eta_{ab}^{l}\sigma_{ij}^{ab}
=\frac{1}{6\pi^{2}}\epsilon_{ijl}(k_{F}^{H}-k_{F}^{L}),
\label{j-Kubo}
\end{equation}
where we used (\ref{eta-eta}) in Appendix \ref{appendix-matrix}.
In contrast to the result (\ref{kubo-nonconserved}) of the
non-conserved spin current, the conductivity for the conserved
spin current (\ref{j-Kubo}) vanishes in the d.c. limit when the
spin-orbit coupling $\gamma_2$ vanishes.

\subsection{Spectral representation of the response function
in terms of the non-abelian gauge field} The Kubo formula result
for the conserved spin current obtained in the previous section
can also be obtained by the spectral representation of the
response function in terms of the eigenstates of the Hamiltonian.
This treatment is similar to the one in quantum Hall effect by
Thouless et al.\cite{thouless1982}. By expressing the Kubo formula
in terms of the eigenstates, we can directly obtain the spin Hall
conductivity in terms of the curvature $F_{ij}$ of the non-abelian
gauge field for each band.

Inserting a set of complete eigenstates into
(\ref{conserved-Kubo}), we obtain
\begin{eqnarray}
&&Q_{ij}^{ab}(i\nu_{m})= -\frac{1}{V}
\sum_{\alpha,\beta,\mathbf{k}} \left( \frac{\langle \alpha
L\mathbf{k}|J_{i}^{ab}|\beta H\mathbf{k}\rangle \langle \beta
H\mathbf{k}|J_{j}|\alpha L\mathbf{k}\rangle }{2\gamma_{2}d/m
+i\nu_{m}} \right.
\nonumber \\
&& \ \ \ \ \ \left. - \frac{\langle \beta
H\mathbf{k}|J_{i}^{ab}|\alpha L\mathbf{k}\rangle \langle \alpha
L\mathbf{k}|J_{j}|\beta H\mathbf{k}\rangle }{-2\gamma_{2}d/m+i\nu_{m}}
\right)(n_{H}-n_{L}),
\end{eqnarray}
where $|\alpha L\mathbf{k}\rangle$ and $|\beta H\mathbf{k}\rangle$
$(\alpha=1,2,\ \beta=1,2)$ are the periodic part of the Bloch
wavefunction with wavenumver $\mathbf{k}$ in the LH and the HH bands,
respectively. By substituting
\begin{equation}
J_{j}=\frac{\partial H}{\partial k_{j}},\ \ J_{i}^{ab}=
\frac{1}{2}\left\{ \frac{\partial H}{\partial k_{i}},\
P^{L}\Gamma^{ab}P^{L} +P^{H}\Gamma^{ab}P^{H} \right\},
\end{equation}
we get
\begin{eqnarray}
&&Q_{ij}^{ab}(i\nu_{m})=\frac{-1}{2V}
\sum_{\alpha,\beta,\gamma,\mathbf{k}} (n_{H}-n_{L})\nonumber \\
&&\ \ \ \cdot\left[\left( \frac{ \langle \beta
L\mathbf{k}|\frac{\partial H}{\partial k_{i}}|\gamma H\mathbf{k} \rangle
\langle \gamma H\mathbf{k}|\frac{\partial H}{\partial k_{j}}|\alpha
L\mathbf{k} \rangle} {2\gamma_{2}d/m+i\nu_{m}}
\right.\right.\nonumber \\
&& \ \ \ \left.+\frac{ \langle \beta L\mathbf{k}|\frac{\partial
H}{\partial k_{j}}|\gamma H \mathbf{k}\rangle \langle \gamma
H\mathbf{k}|\frac{\partial H}{\partial k_{i}}|\alpha L\mathbf{k} \rangle}
{2\gamma_{2}d/m-i\nu_{m}}\right) \langle \alpha
L\mathbf{k}|\Gamma^{ab}|\beta L\mathbf{k} \rangle
\nonumber \\
&&\ \ \ +\left.\left( \frac{ \langle \beta H \mathbf{k}|\frac{\partial
H}{\partial k_{j}}|\gamma L\mathbf{k} \rangle \langle \gamma
L\mathbf{k}|\frac{\partial H}{\partial k_{i}}|\alpha H \mathbf{k}\rangle}
{2\gamma_{2}d/m+i\nu_{m}}
\right.\right.\nonumber \\
&&\ \ \  \left. \left.+\frac{ \langle \beta H\mathbf{k}|\frac{\partial H}{\partial
k_{i}}|\gamma L\mathbf{k} \rangle \langle \gamma
L\mathbf{k}|\frac{\partial H}{\partial k_{j}}|\alpha H\mathbf{k} \rangle}
{2\gamma_{2}d/m-i\nu_{m}}\right)\right.\nonumber \\
&&\makebox[2cm]{}\ \ \left.\cdot \langle \alpha
H\mathbf{k}
|\Gamma^{ab}|\beta H \mathbf{k}\rangle\right]. 
\label{Qijab}\end{eqnarray} It can
be checked that $Q_{ij}^{ab}(i\nu_{m}=0)=0$.
Here we shall use the Feynman-Hellman theorem;
because $H|\gamma H\mathbf{k}\rangle=
E_{H}|\gamma H\mathbf{k}\rangle$ implies
\begin{equation}
\frac{\partial H}{\partial k_{i}}|\gamma H\mathbf{k}\rangle
+H\left|\frac{\partial (\gamma H\mathbf{k})}{\partial k_{i}}\right\rangle
=
\frac{\partial E_{H}}{\partial k_{i}}|\gamma H\mathbf{k}\rangle
+E_{H}\left|\frac{\partial (\gamma H\mathbf{k})}{\partial k_{i}}\right\rangle,
\end{equation}
it follows that
\begin{eqnarray}
&&\left\langle\beta L\mathbf{k}\left|\frac{\partial H}{\partial k_{i}}
\right|\gamma H \mathbf{k}\right\rangle=-\frac{2\gamma_{2}d(\mathbf{k})}{m}
\left\langle
\beta L\mathbf{k}\left|\frac{\partial(\gamma H\mathbf{k})}{\partial k_{i}}
\right\rangle\right.\nonumber \\
&&\ \ \  =
\frac{2\gamma_{2}d(\mathbf{k})}{m}
\left.\left\langle\frac{\partial (\beta L\mathbf{k})}{\partial
k_{i}}\right|\gamma H\mathbf{k}\right\rangle.
\end{eqnarray}
Therefore, in the d.c. limit
\begin{eqnarray}
&&\sigma_{ij}^{ab}=\lim_{\omega\rightarrow 0}
\frac{Q_{ij}^{ab}(\omega)}{-i\omega}=-\frac{i}{2V}
\sum (n_{H}-n_{L})\nonumber \\
&&\ \ \ \cdot\left[\left( -\left\langle \frac{\partial (\beta
L\mathbf{k})}{\partial k_{i}}\right| P^{H}\left| \frac{\partial
(\alpha L\mathbf{k})}{\partial k_{j}}\right\rangle \right.\right.
\nonumber \\
&&\makebox[1cm]{}\left.
+\left\langle
\frac{\partial (\beta L\mathbf{k})}{\partial k_{j}}\right
|P^{H}\left| \frac{\partial (\alpha L\mathbf{k})}{\partial
k_{i}}\right\rangle\right)
\nonumber \\
&& \makebox[2.5cm]{} \cdot\langle \alpha
L\mathbf{k}|\Gamma^{ab}|\beta L\mathbf{k} \rangle
\nonumber \\
&&\ \ \ +\left( -\left\langle \frac{\partial (\beta
H\mathbf{k})}{\partial k_{j}}\right|P^{L}\left| \frac{\partial
(\alpha H\mathbf{k})}{\partial k_{i}}\right\rangle \right.
\nonumber \\
&&\makebox[1cm]{}
+\left.\left\langle
\frac{\partial (\beta H\mathbf{k})}{\partial
k_{i}}\right|P^{L}\left| \frac{\partial (\alpha
H\mathbf{k})}{\partial k_{j}}\right\rangle
\right)\nonumber \\
&&\makebox[2.5cm]{} \left.\cdot\langle \alpha
H\mathbf{k}|\Gamma^{ab}|\beta H\mathbf{k} \rangle \right]
\end{eqnarray}
This formula can be expressed with the field strength $F_{ij}$ of
the SU(2) gauge field for each band. We define the gauge field for
the LH band as
\begin{equation}
(A^{L}_{i})_{\alpha\beta}=-i\left\langle \alpha L\mathbf{k}
\left|\frac{\partial}{\partial k_{i}} \right|\beta L\mathbf{k}\right\rangle ,
\end{equation}
and similarly for $A^{H}_{i}$. The corresponding field strength is
\begin{equation}
F^{L}_{ij}= \frac{\partial A^{L}_{j}}{\partial k_{i}}
-\frac{\partial A^{L}_{i}}{\partial k_{j}}
+i[A^{L}_{i},A^{L}_{j}],
\end{equation}
and $F_{ij}^{H}$, respectively. While in this definition
$A^{L}_{i}$ is a $2\times 2$ matrix, it can be embedded into
$4\times 4$ matrix by identifying it with $|\alpha L
\mathbf{k}\rangle (A^{L}_{i})_{\alpha\beta}\langle \beta
L\mathbf{k}|$. We use the same notation $A^{L}_{i}$ to denote the
$4\times 4$ matrix defined in this way. The $4\times 4$ matrices
$A^{H}_{i}$, $F^{L}_{ij}$, and $F^{H}_{ij}$ are defined similarly.
They can be expressed as linear combinations of $\Gamma^{ab}$ as
\begin{equation}
F^{L}_{ij}=F^{L,ab}_{ij}\Gamma^{ab},\
F^{H}_{ij}=F^{H,ab}_{ij}\Gamma^{ab},
\end{equation}
Then the resulting form of the spin Hall conductivity is obtained
as
\begin{equation}
\sigma_{ij}^{ab}=\frac{4}{V} \sum_{\mathbf{k}}(n_{H}-n_{L}) \left(
-F^{L,ab}_{ij}+F^{H,ab}_{ij}\right),
\end{equation}
in exact agreement with (\ref{sigmaijab}).
By
contracting with $\frac{1}{3}\eta^{l}_{ab}$ as in (\ref{j-Kubo}),
we get
\begin{eqnarray}
&&\sigma^{l}_{ij\text{(c)}}
\equiv\frac{1}{3}\eta_{ab}^{l}\sigma_{ij}^{ab} \nonumber \\
&&\ \ \ = \frac{4}{3V}\eta_{ab}^{l} \sum_{\mathbf{k}}(n_{H}-n_{L}) \left(
-F^{L,ab}_{ij} +F_{ij}^{H,ab}
\right)\nonumber \\
&&\ \ \ =\frac{1}{6\pi^{2}}\epsilon_{ijl}(k_{F}^{H}-k_{F}^{L}),
\label{prb-result}
\end{eqnarray}
in exact agreement with (\ref{j-Kubo}).

\subsection{Semiclassical limit}
The above result can be written as correlation functions in a
real-time formalism;
\begin{equation}
{\sigma}^{l}_{ij\text{(c)}}= \frac{1}{6\omega
Z}\text{tr}\int_{0}^{\infty}dt\ e^{i(\omega+i\delta)t}
[\{J_{i}(t),\ S^{l}_{\text{(c)}}\},\ J_{j}]e^{-\beta H},
\label{quantum-sigma}
\end{equation}
where $Z=\text{tr}e^{-\beta H}$ is the
partition function of the equilibrium. This quantity does not
change if we replace $S^{l}_{\text{(c)}}$ defined in (\ref{sk-c}) by
$S^{\prime l}_{\text{(c)}}=\lambda \hat{k}^{l}$, which follows from
the fact that the helicity is a conserved quantum number.

In a semiclassical (sc) approximation, one treats the spin
$S^{\prime l}_{\text{(c)}}$ as a classical variable, commuting with
the current $J_{j}$. Under this approximation, one obtains
\begin{equation}
\sigma^{l}_{ij\text{(c)}} \text{(sc)} = \frac{1}{3\omega
Z}\text{tr}\int_{0}^{\infty}dt\ e^{i(\omega+i\delta)t} [J_{i}(t),\
J_{j}]S^{\prime l}_{\text{(c)}}e^{-\beta H}, \label{sigmaijkapprox}
\end{equation}
where we used the fact that $S^{\prime l}_{ \text{(c)}}$ commutes
with $H$. Direct computation of this correlation function leads to
the semiclassical result
\begin{eqnarray}
&&\sigma^{l}_{ij\text{(c)}} \text{(sc)} \nonumber \\
&&\ \ \ = \frac{1}{3V} \sum_{\mathbf{k}}\left( n_{H}{\rm tr}
F_{ij}^{H}P^{H}S^{l}P^{H} +n_{L}{\rm tr} F_{ij}^{L}P^{L}S^{l}P^{L}
\right)\nonumber\\
&&\ \ \ =\frac{1}{12\pi^{2}}\epsilon_{ijl}(3k_{F}^{H}-k_{F}^{L}).
\label{Science-result}
\end{eqnarray}
which agrees exactly with the semiclassical results
\cite{murakami2003} based on the wave-packet equation of motion.
The noncommutativity between the quantum spin and current
operators contained in (\ref{quantum-sigma}) leads to a quantum
correction
\begin{eqnarray}
\Delta\sigma^{l}_{ij\text{(c)}}
=\sigma^{l}_{ij\text{(c)}}-\sigma^{l}_{ij\text{(c)}}\text{(sc)}
=-\frac{1}{12\pi^{2}}\epsilon_{ijl}(k_{F}^{H}+k_{F}^{L}).
\label{difference}
\end{eqnarray}
to the semiclassical result (\ref{Science-result}).

We would like to stress that this difference arises from the
definition of spin current. In (\ref{prb-result}), we defined the
spin current as an anticommutator between velocity and the spin as
(\ref{anticommutator}). This definition of spin current amounts to
taking the spin as a quantum average between the initial state and
the intermediate state in the Kubo formula, as can be seen from
Eq.~(\ref{Qijab}). On the other hand, the semiclassical result
\cite{murakami2003} corresponds to taking the spin as that of the
initial state. In this semiclassical formalism, the wave-packets with different
helicities have the opposite transverse velocities with respect to
the external electric field.

\section{Conclusions and Discussions}
\label{conclusion} In the present paper, we studied the spin Hall
effect in hole-doped semiconductors such as Ge and GaAs.
The four valence bands, which are made 
out of p-orbitals with the spin-orbit interaction, consists 
of the doubly
degenerate heavy-hole band and light-hole band. (When we assume
the inversion symmetry, the Kramers theorem requires at least
double degeneracy at each $\mathbf{k}$-point.) 
These two bands touch at the $\Gamma$-point.
The effective Hamiltonian
describing these valence bands, so-called the Luttinger Hamiltonian,
has a beautiful mathematical structure described by the SO(5) 
Clifford algebra. At a given momentum $\mathbf{k}$, the spin-orbit coupling
singles out a fixed direction in the five-dimensional space of the 
$\mathbf{d}$
vectors, and breaks the symmetry down to
SO(4)=SU(2)$\times$SU(2). This symmetry property can be used to
define conserved spin currents in both the LH and the HH bands.
The quantum response of the conserved spin current can be
calculated exactly within the Kubo formalism, and the result is
summarized in Eq.~(\ref{prb-result}). This result can be expressed
in terms of purely geometric quantities, or equivalently, in terms
of the non-abelian Yang monopole field strength, defined in the
five-dimensional space of the $\mathbf{d}$ vectors. This result also
establishes the deep connection between the spin current in the
Luttinger model and the 4DQHE model of Zhang and
Hu\cite{zhang2001}, which also uses the Yang monopole as the
non-abelian background gauge field. In the former case, the Yang
monopole is defined in momentum space over the space of the five-dimensional 
$\mathbf{d}$ vectors, while in the latter case, the Yang
monopole is defined in the real space. Magnetic monopole structure
in the five dimensional momentum space has also been discussed by
Volovik\cite{volovik2001}.

Our fully quantum mechanical results are compared with previous 
semiclassical one (\ref{Science-result}), 
and a quantum correction due to the entanglement
of spin and velocity is identified.
The quantum correction can be traced to the non-commutativity 
and entanglement between the spin and the current operator. 
In physical systems where this entanglement is destoyed
by some decoherence mechanisms, the semiclassical result
might be realized.
In Ref.\ \onlinecite{culcer2003}, Culcer et al.\ developed a wavepacket 
formalism, and discussed the difference between our semiclassical 
result \cite{murakami2003} and the Kubo-formula result 
(\ref{kubo-nonconserved}) 
using the conventional definition of the spin current.
They incorporated the nonzero correlation between spin and velocity 
into a ``spin dipole'' and 
``torque moment'' terms in their wavepacket formalism, and
reproduced the Kubo-formula result Eq.\ (\ref{kubo-nonconserved}) after also
including a first-order 
field correction to the wavepacket spin.

In the calculations of the spin current presented in this paper,
we assumed an absence of impurities. On the other hand,
we have also done a calculation 
including a scattering by randomly-distributed impurities.
By assuming that the scattering potential is isotropic and
accompanies no spin-flip, we calculated the spin 
current within the Born approximation and the ladder approximation
for the vertex correction. The self-energy obtains a finite imaginary
part $\hbar/2\tau$ as usual, where $\tau$ is a lifetime.
The vertex correction, on the other hand, vanishes due to the
parity, namely because the Hamiltonian is an even function
of $\mathbf{k}$. Thus as far as the broadening of the energy
$\hbar/\tau$ is much smaller than the energy difference between
two bands $E_{L}-E_{H}$, the spin current calculated in
(\ref{j-Kubo}) remains unchanged.
The details of the calculation are involved and 
will be presented elsewhere.

The dissipationless spin current discovered in recent theoretical
works has many profound consequences both in fundamental science
and in technological applications. However, in models investigated
so far, there is still a finite longitudinal charge conductivity
and dissipation associated with charge transport. A key objective
along the current line of research is to identify spin-orbit
coupled system with a gap in the electronic excitation spectrum,
which might lead to quantized spin Hall effect, similar to the
familiar quantized Hall effect. This exciting possibility is
suggested by the fact that $\sigma_{ij}^{l}$ is represented as the
integral of the gauge curvature over the occupied states, and does
not require the Fermi surface across which the particle-hole
excitation occurs.

\appendix

\section{$\Gamma$ matrices and related identities}
\label{appendix-matrix} With the expressions for the $\mathbf{S}$
matrices
{\arraycolsep=2pt
\begin{eqnarray}
&&S^{z}=\left(
\begin{array}{cccc}
\frac{3}{2}&&&\\
&\frac{1}{2}&&\\
&&-\frac{1}{2}&\\
&&&-\frac{3}{2}
\end{array}
\right),\ S^{x}=\left(
\begin{array}{cccc}
&\frac{\sqrt{3}}{2}&&\\
\frac{\sqrt{3}}{2}&&1&\\
&1&&\frac{\sqrt{3}}{2}\\
&&\frac{\sqrt{3}}{2}&
\end{array}
\right),\nonumber \\
&& S^{y}=\left(
\begin{array}{cccc}
&-\frac{\sqrt{3}}{2}i&&\\
\frac{\sqrt{3}}{2}i&&-i&\\
&i&&-\frac{\sqrt{3}}{2}i\\
&&\frac{\sqrt{3}}{2}i&
\end{array}
\right),\ \label{jxyz}
\end{eqnarray}
}we get
\begin{eqnarray}
&&S^{x2}=\frac{\sqrt{3}}{2}\sigma^{x}\otimes 1 -
\frac{1}{2}\sigma^{z}\otimes\sigma^{z}+\frac{5}{4},\\
&&S^{y2}=-\frac{\sqrt{3}}{2}\sigma^{x}\otimes 1 -
\frac{1}{2}\sigma^{z}\otimes\sigma^{z}+\frac{5}{4},\\
&&S^{z2}=\sigma^{z}\otimes\sigma^{z}+\frac{5}{4},
\end{eqnarray}
\begin{eqnarray}
&&S^{x}S^{y}+S^{y}S^{x}=\sqrt{3}\sigma^{y}\otimes 1 ,\\
&&S^{y}S^{z}+S^{z}S^{y}=\sqrt{3}\sigma^{z}\otimes \sigma^{y} ,\\
&&S^{z}S^{x}+S^{x}S^{z}=\sqrt{3}\sigma^{z}\otimes \sigma^{x},
\end{eqnarray}
where $\sigma^{i}$ $(i=1,2,3)$ are the Pauli matrices.
Let us define the $\Gamma$ matrices as
\begin{eqnarray}
&&\Gamma^{1}=\sigma^{z}\otimes\sigma^{y}= \frac{1}{\sqrt{3}}
(S^{y}S^{z}+S^{z}S^{y}),
\\
&&\Gamma^{2}=\sigma^{z}\otimes\sigma^{x} =\frac{1}{\sqrt{3}}
(S^{z}S^{x}+S^{x}S^{z}),\\
&&\Gamma^{3}=\sigma^{y}\otimes 1 =\frac{1}{\sqrt{3}}
(S^{x}S^{y}+S^{y}S^{x}), \\
&&\Gamma^{4}=\sigma^{x}\otimes 1=\frac{1}{\sqrt{3}}
(S^{x2}-S^{y2}),\\
&&\Gamma^{5}=\sigma^{z}\otimes\sigma^{z}=S^{z2}-\frac{5}{4}.
\end{eqnarray}

Since $\Gamma^{a}\Gamma^{b}+\Gamma^{b}\Gamma^{a}=2\delta_{ab}$,
These five matrices generate the SO(5) Clifford algebra
\cite{georgi}. We shall define the traceless symmetric tensor
$\xi_{a}^{ij}$ by (\ref{gamma}), i.e.
\begin{equation}
\Gamma^{a}=\xi_{a}^{ij}\{S^{i},\ S^{j}\},\ \xi_{a}^{ij}=\xi_{a}^{ji},\
\xi_{a}^{ii}=0.
\end{equation}
Explicitly they are written as
\begin{eqnarray}
&&\xi_{1}^{yz}=\frac{1}{2\sqrt{3}},\
\xi_{2}^{zx}=\frac{1}{2\sqrt{3}},\
\xi_{3}^{xy}=\frac{1}{2\sqrt{3}},\nonumber \\
&&\xi_{4}^{xx}=-\xi_{4}^{yy}=\frac{1}{2\sqrt{3}},\nonumber \\
&&\xi_{5}^{xx}=\xi_{5}^{yy}=-\frac{1}{6},\
\xi_{5}^{zz}=\frac{1}{3},\nonumber
\end{eqnarray}
and those obtained by $\xi^{ij}_{a}=\xi^{ji}_{a}$.
They form the vector representation of the SO(5) algebra, and are
expressed as $4\times 4$ Hermitian matices. When we define a
representation in this space of $4\times 4$ Hermitian matrices as
$\Gamma^{ab}|A\rangle =|[\Gamma^{ab}, A]\rangle$, $A^{\dagger}=A$,
$\Gamma^{ab}=\frac{1}{2i}[\Gamma^{a}, \Gamma^{b}]$, it is shown to
be a product of two four-dimensional spinor representations of
SO(5). This product of two spinor representations
can be classified into the irreducible representations of SO(5),
and each irreducible representation is expressed as a product of
the elements of the Clifford algebra. Thus
$\bf{4}\times\bf{4}=\bf{1}+\bf{5}+\bf{10}$, where $\bf{4}$ is the
spinor representation, $\bf{1}$ is a trivial representation,
$\bf{5}$ is a vector representation spanned by $\Gamma^{a}$, and
$\bf{10}$ is an adjoint representation spanned by $\Gamma^{ab}$.
These matrices $1$, $\Gamma^{a}$, and $\Gamma^{ab}$, span the
space of 4$\times$4 Hermitian matrices. Moreover, because
$\Gamma^{1}\Gamma^{2}\Gamma^{3}\Gamma^{4}\Gamma^{5}=-1$, a product
of more than two $\Gamma$ matrices can be written as a linear
combination of $1$, $\Gamma^{a}$, and $\Gamma^{ab}$. It is thus
possible to write $S^{i}$ in terms of these matrices as
\begin{eqnarray}
&&S^{x}=\frac{\sqrt{3}}{2}1\otimes\sigma^{x}+\frac{1}{2}
(\sigma^{x}\otimes\sigma^{x}+\sigma^{y}\otimes\sigma^{y})
\nonumber \\
&&\ \ \ =\frac{\sqrt{3}}{2}\Gamma^{15} -\frac{1}{2}
(\Gamma^{23}-\Gamma^{14}),\label{jx}\\
&&S^{y}=\frac{\sqrt{3}}{2}1\otimes\sigma^{y}+\frac{1}{2}
(-\sigma^{x}\otimes\sigma^{y}+\sigma^{y}\otimes\sigma^{x})\nonumber \\
&&\ \ \ =-\frac{\sqrt{3}}{2}\Gamma^{25} +\frac{1}{2}
(\Gamma^{13}+\Gamma^{24}),\label{jy}\\
&&S^{z}=\sigma^{z}\otimes 1+\frac{1}{2}1\otimes\sigma^{z}
=-\Gamma^{34} -\frac{1}{2} \Gamma^{12}
\label{jz}.
\end{eqnarray}
These are used to calculate the correlation function in the Kubo
formula. To formulate the problem in a covariant fashion, we
define $\eta_{ab}^{i}$ as $ S^i = \eta^i_{ab} \Gamma^{ab}$, where
$\Gamma^{ab}=\frac{1}{2i}[\Gamma^{a},\Gamma^{b}]$ are generators
of the SO(5) algebra, and $\eta^{i}_{ab}=-\eta^{i}_{ba}$. Nonzero
components of $\eta^{i}_{ab}$ are
\begin{eqnarray*}
&&\eta_{15}^{x}=\frac{\sqrt{3}}{4}, \ \eta_{23}^{x}=-\frac{1}{4},
\
\eta_{14}^{x}=\frac{1}{4}, \\
&&\eta_{25}^{y}=-\frac{\sqrt{3}}{4}, \ \eta_{13}^{y}=\frac{1}{4},
\
\eta_{24}^{y}=\frac{1}{4}, \\
&& \eta_{34}^{z}=-\frac{1}{2}, \ \eta_{12}^{z}=-\frac{1}{4},
\end{eqnarray*}
and the ones obtained by $\eta^{i}_{ab}=-\eta^{i}_{ba}$.
The ten $\Gamma^{ab}$ matrices contain both  the three spin operators
$S^i$ and seven cubic, symmetric and traceless combinations of
the spin operators
of the form $S^i S^j S^k$. These seven cubic operators are
\begin{eqnarray}
&&(S^x )^3=\frac{7\sqrt{3}}{8}\Gamma^{15}+\frac{7}{8}\Gamma^{14}-\frac{13}{8}
\Gamma^{23},\\
&&(S^y )^3=-\frac{7\sqrt{3}}{8}\Gamma^{25}+\frac{7}{8}\Gamma^{24}+\frac{13}{8}
\Gamma^{13},\\
&&(S^z)^3=-\frac{13}{8}\Gamma^{12}-\frac{7}{4}\Gamma^{34},\\
&&\{ S^x, (S^y)^2-(S^z)^2\}
=-\frac{\sqrt{3}}{2}\Gamma^{15}+\frac{3}{2}\Gamma^{14},\\
&&\{ S^y, (S^z)^2-(S^x)^2\}
=-\frac{\sqrt{3}}{2}\Gamma^{25}-\frac{3}{2}\Gamma^{24},\\
&&\{ S^z, (S^x)^2-(S^y)^2\}
=\sqrt{3}\Gamma^{35},\\
&&S^x S^y S^z +S^z S^y S^x =-\frac{\sqrt{3}}{2}\Gamma^{45}.
\end{eqnarray}

There are several useful formulae for $\Gamma^{a}$, which are used
in the calculation in this paper:
\begin{eqnarray}
&&[\Gamma^{ab},\Gamma^{c}]=2i(\delta_{ac}\Gamma^{b}-\delta_{bc}\Gamma^{a}),\\
&&\{\Gamma^{ab},\Gamma^{c}\}
=\epsilon_{abcde}\Gamma^{de}, \label{GammaId} \\
&&[\Gamma^{ab},\Gamma^{cd}]=-2i(\delta_{bc}\Gamma^{ad}-\delta_{bd}\Gamma^{ac}
\nonumber \\
&&\makebox[3cm]{}
-\delta_{ac}\Gamma^{bd}+\delta_{ad}\Gamma^{bc}), \\
&&\{\Gamma^{ab},\Gamma^{cd}\}=
2\epsilon_{abcde}\Gamma^{e}+2\delta_{ac}\delta_{bd}-2
\delta_{ad}\delta_{bc}, \label{Gammaab-Gammacd}
\end{eqnarray}
\begin{eqnarray}
&&\text{tr}(\Gamma^{a}\Gamma^{b})=4\delta_{ab},\\
&&\text{tr}(\Gamma^{a}\Gamma^{b}\Gamma^{c})=0,\\
&&\text{tr}(\Gamma^{a}\Gamma^{b}\Gamma^{c}\Gamma^{d})=4
(\delta_{ab}\delta_{cd}+\delta_{ad}\delta_{bc}-\delta_{ac}\delta_{bd}),\\
&&\text{tr}(\Gamma^{a}\Gamma^{b}\Gamma^{c}\Gamma^{d}\Gamma^{e})=
-4\epsilon_{abcde}.
\end{eqnarray}

By substituting $S^{i}=\eta^{i}_{ab}\Gamma^{ab}$ into
the commutation relation $[S^{i},S^{j}]=i\epsilon_{ijk}S^{k}$,
one can easily derive
\begin{equation}
[\eta^{i},\ \eta^{j}]=-\frac{1}{4}\epsilon_{ijk}\eta^{k},
\label{eta-algebra}
\end{equation}
where
${\eta}^{i}$ is a 5$\times$5 matrix with components $\eta_{ab}^{i}$.
In other words, the matrices $-4i\eta^{i}$ form the spin-2
representation of the SU(2) algebra. It can also be shown that
\begin{equation}
\eta_{ab}^{i}\eta_{ab}^{j}=-\text{tr}(\eta^{i}\eta^{j})=\frac{5}{8}\delta_{ij},
\label{eta-eta}
\end{equation}
and
\begin{equation}
\eta^{i}\eta^{i}=-\frac{3}{8}.
\label{etai-etai}
\end{equation}

Let us write down the formula for $d_{a}$.
We can easily check that
\begin{equation}
\xi_{a}^{ij}\xi_{a}^{kl}
=\frac{1}{12}(\delta_{ik}\delta_{jl}+\delta_{il}\delta_{jk})-\frac{1}{18}
\delta_{ij}\delta_{kl}.
\end{equation}
Then it follows that
\begin{eqnarray}
&&(\xi_{a}^{ij}k_{i}k_{j})\Gamma^{a}=\xi_{a}^{ij}\xi_{a}^{kl}k_{i}k_{j}\{
S^{k},\ S^{l}\}\nonumber \\
&&\ \ = \frac{1}{3}(\mathbf{k}\cdot\mathbf{S})^{2}-\frac{5}{12}k^{2}.
\end{eqnarray}
Therefore, by substituting
\begin{equation}
(\mathbf{k}\cdot\mathbf{S})^{2}=\frac{5}{4}k^{2}+3\xi_{a}^{ij}k_{i}k_{j}
\Gamma^{a}.
\label{kS}
\end{equation}
into the Luttinger Hamiltonian (\ref{Luttinger}) and comparing it with
(\ref{Hamiltonian}), we get
\begin{equation}
d_{a}=-3\xi_{a}^{ij}k_{i}k_{j},
\end{equation}
in accordance with (\ref{d}).
This tensor $\xi^{ij}_{a}$ can be expressed
in terms of $\eta_{cd}^{k}$ as calculated below.
\begin{eqnarray}
&&(\mathbf{k}\cdot\mathbf{S})^{2}
=\frac{1}{2}\left\{k_{i}\eta_{ab}^{i}\Gamma^{ab},
\ k_{j}\eta_{cd}^{j}\Gamma^{cd}\right\}\nonumber \\
&&\ \  \ \ =\frac{5}{4}k^{2}+k_{i}k_{j}\epsilon_{abcde}
\eta_{ab}^{i}\eta_{cd}^{j}\Gamma^{e}.
\end{eqnarray}
By comparing with Eq.~(\ref{kS}) we get
\begin{equation}
\xi^{ij}_{e}=\frac{1}{3} \epsilon_{abcde}\eta_{ab}^{i}\eta_{cd}^{j}.
\end{equation}
One can also check that
\begin{equation}
\xi_{a}^{ij}\xi_{b}^{ij}=\frac{1}{6}\delta_{ab}.
\end{equation}

\section{Details of the Kubo formula calculations}
\label{appendix-kubo} The electron Green's function is written as
\begin{eqnarray}
&&G_{\mu\nu}(\mathbf{k},i\omega_{n})=
\left(\frac{1}{i\omega_{n}-H+\mu}\right)_{\mu\nu}\nonumber \\
&& \ \ \
=\frac{1}{(i\omega_{n}+\mu-\epsilon(\mathbf{k}))^{2}-\gamma_{2}^{2}d^{2}/m^{2}
}\nonumber
\\&&\ \ \ \ \ \ \ \cdot \left(
i\omega_{n}+\mu-\epsilon({\bm
k})+\frac{\gamma_{2}}{m}d_{a}(\mathbf{k})
\Gamma^{a}\right)_{\mu\nu} \nonumber
\\
&&\ \ \ =f(k,i\omega_{n})\left(g(k,i\omega_{n})+
\frac{\gamma_{2}}{m}d_{a}(\mathbf{k})\Gamma^{a}\right)_{\mu\nu}.
\label{Green}
\end{eqnarray}
In the clean limit, the Kubo formula calculation proceeds as
follows
\begin{eqnarray}
&&Q_{ij}^{ab}(i\nu_{m})= -\frac{1}{V}\int_{0}^{\beta} \langle \hat{T}
J_{i}^{ab}(u){J}_{j}\rangle e^{i\nu_{m}u}du
\nonumber \\
&&\ \ \ =\frac{1}{V\beta}\sum_{k,n} \text{tr}\left(
J_{i}^{ab}G(\mathbf{k},i(\omega_{n}+\nu_{m}))J_{j} G
(\mathbf{k},i\omega_{n})\right)\nonumber \\
&&\ \ \ =\frac{1}{V\beta}\sum_{k,n} f\left(k,
i(\omega_{n}+\nu_{m})\right) f\left(k, i\omega_{n}\right)
\nonumber \\
&&\ \ \ \ \ \ \cdot \text{tr}\left[ \left( \frac{\partial
\epsilon}{\partial k_{i}} P_{ab,cd}\Gamma^{cd} +\frac{1}{2}
\frac{\gamma_{2}}{m}\frac{\partial d_{f}}{\partial k_{i}}
 P_{ab,cd}
\epsilon_{fcdmn}\Gamma^{mn} \right)\right.
\nonumber \\
&&\ \ \ \ \ \ \cdot \left.\left\{ g\left( k,
i(\omega_{n}+\nu_{m})\right)+\frac{\gamma_{2}}{m}d_{g}
\Gamma^{g}\right\} \right.
\nonumber \\
&&\ \ \ \ \ \ \cdot \left. \left( \frac{\partial
\epsilon}{\partial k_j}+ \frac{\gamma_{2}}{m}\frac{\partial
d_{h}}{\partial k_{j}} \Gamma^{h}\right)\left\{ g(k,i\omega_{n})+
\frac{\gamma_{2}}{m}d_{l}\Gamma^{l}\right\} \right], \label{Qij}
\end{eqnarray}
where we used (\ref{GammaId}).

To evaluate the summation over $\omega_{n}$, we use a formula
\begin{eqnarray}
&&\frac{1}{\beta}\sum_{n}f(k, i(\omega_{n}+\nu_{m}))f(k,
i\omega_{n})
(Cg(k,i\omega_{n})+D)\nonumber \\
&&\ \ \
=\frac{m}{\gamma_{2}}\frac{\left(\frac{-i\nu_{m}}{2}C+D\right)
(n_{F}(\epsilon_{L})-n_{F}(\epsilon_{H})) }{d(\mathbf{k})
\left((i\nu_{m})^{2}-4\gamma_{2}^{2}d(\mathbf{k})^{2}/m^{2}\right)},
\end{eqnarray}
where $C,D$ are constants. By noting that the term proportional to
$g(k,i\omega_{n}+i\nu_{m})g(k,i\omega_{n})$ becomes zero in taking
the trace of the matrix, we have,
\begin{eqnarray}
&&Q_{ij}^{ab}(i\nu_{m})\nonumber \\
&&\ \ \ =\frac{1}{V}\sum_{\mathbf{k}} \frac{m}{\gamma_{2}}
\frac{n_{F}(\epsilon_{L})-n_{F}(\epsilon_{H})}{d((i\nu_m)^{2}-4\gamma_{2}^{2}
d^{2}/m^{2})}
\nonumber \\
&&\ \ \ \ \ \ \cdot \text{tr}\left[ \left( \frac{\partial
\epsilon}{\partial k_{i}} P_{ab,cd}\Gamma^{cd}
+\frac{1}{2}\frac{\gamma_{2}}{m} \frac{\partial d_{f}}{\partial
k_{i}}
 P_{ab,cd}
\epsilon_{fcdmn}\Gamma^{mn}
\right)\right.\nonumber \\
&&\ \ \ \ \ \ \cdot \left.\left(
\frac{i\nu_{m}}{2}+\frac{\gamma_{2}}{m} d_{g} \Gamma^{g}\right)
\right.
\nonumber \\
&&\ \ \ \ \ \ \cdot \left. \left( \frac{\partial
\epsilon}{\partial k_j} +\frac{\gamma_{2}}{m}\frac{\partial
d_{h}}{\partial k_{j}} \Gamma^{h}\right) \left(
-\frac{i\nu_{m}}{2} +\frac{\gamma_{2}}{m}d_{l}\Gamma^{l}\right)
\right].
\end{eqnarray}
The matrix inside the trace is a linear combination of products of
two, three, four and five $\Gamma$ matrices. By taking the trace,
only the products of four and five $\Gamma$ matrices survive. It
is worth noting that the $\frac{\partial\epsilon}{\partial k_{i}}$
and $\frac{\partial\epsilon}{\partial k_{j}}$ gives no
contribution; the former is because of $ P_{ab,cd}d_{d}=0$ and $
\epsilon_{cdghl}d_{g}d_{l}=0$, and the
latter is due to $d_{g}\Gamma^{g}d_{l}\Gamma^{l}=d^{2}$. After
some calculation it becomes,
\begin{equation}
Q_{ij}^{ab}(i\nu_{m})=\frac{-16\nu_{m}}{V}
\left(\frac{\gamma_{2}}{m}\right)^{2}\sum_{\mathbf{k}}
\frac{n_{F}(\epsilon_{L})-n_{F}(\epsilon_{H})}{
(i\nu_m)^{2}-4\gamma_{2}^{2}d^{2}/m^{2}}d^{2}
G^{ab}_{ij}.
\end{equation}
In the d.c. limit we have,
\begin{eqnarray}
&&\sigma_{ij}^{ab} = \lim_{\omega\rightarrow 0}
\frac{Q_{ij}^{ab}(\omega)}{-i\omega}
=\frac{4}{V}\sum_{\mathbf{k}}
(n_{L}(\mathbf{k})-n_{H}(\mathbf{k}))
G_{ij}^{ab}\nonumber \\
&&=\frac{2}{V}\sum_{\mathbf{k}}\frac{\epsilon_{ijl}k_{l}}{k^{6}}
(16(\mathbf{k}\cdot\bm{\eta})^{3}+k^{2}
\mathbf{k}\cdot\bm{\eta})_{ab} (n_{L}-n_{H}),\ \ \ 
\end{eqnarray}
where we substituted (\ref{Gabij}).
Because of the spherical symmetry of the problem,
the summation over $\mathbf{k}$ can be simplified further.
By using identities
\begin{eqnarray}
&&\sum_{\mathbf{k}}\Phi(k)k_{i}k_{j}=\frac{1}{3}\delta_{ij}
\sum_{\mathbf{k}}\Phi(k)k^{2},\\
&&\sum_{\mathbf{k}}\Phi(k)k_{i}k_{j}k_{k}k_{l}\nonumber \\
&&\ \ \ \ \ \
=\frac{1}{15}(\delta_{ij}\delta_{kl}+
\delta_{ik}\delta_{jl}+\delta_{il}\delta_{jk})\sum_{\mathbf{k}}\Phi(k)k^{4},
\end{eqnarray}
where $\Phi(k)$ is an arbitrary function of $k=|\mathbf{k}|$,
we can calculate as
\begin{eqnarray}
&&\sum_{k}\frac{1}{k^4}k_{l}(\mathbf{k}\cdot\bm{\eta})(n_{L}-n_{H})
=\frac{1}{3}\sum_{k}\frac{n_{L}-n_{H}}{k^2}\eta^{l}\\
&&\sum_{k}\frac{1}{k^6}k_{l}(\mathbf{k}\cdot\bm{\eta})^{3}
(n_{L}-n_{H}),\nonumber \\
&&\ \ \
=\frac{1}{15}\sum_{k}\frac{1}{k^2}(\eta^{i}\eta^{i}\eta^{l}
+\eta^{i}\eta^{l}\eta^{i}+\eta^{l}\eta^{i}\eta^{i})
(n_{L}-n_{H})\nonumber\\
&&\ \ \ =-\frac{17}{240}\sum_{k}\frac{n_{L}-n_{H}}{k^2}\eta^{l},
\end{eqnarray}
where we used (\ref{eta-algebra}) (\ref{etai-etai}).
Hence
\begin{equation}
\sigma_{ij}^{ab} =-\frac{8}{5V}\eta_{ab}^{l}\epsilon_{ijl}
\sum_{k}\frac{n_{L}-n_{H}}{k^{2}}
=-\frac{4}{5\pi^{2}}\eta_{ab}^{l}\epsilon_{ijl}(k_{F}^{L}-k_{F}^{H}).
\end{equation}

\section{Magnetic monopoles in $d=3$ and $d=5$}
\label{appendix-gauge} From Eq.~(\ref{Hamiltonian}) we see that
the microscopic Hamiltonian depends on $\mathbf{k}$ only through the 5D
vector $\mathbf{d}(\mathbf{k})$;
therefore, it is natural to define the most general
5D gauge connection in the $\mathbf{d}$ space, and then project the gauge
connection to the 3D $\mathbf{k}$ space. Let $P^{L}$ and $P^{H}$ the
projections onto the LH and HH bands. These projections
 have the following properties;
\begin{eqnarray*}
&&P^{L}=\frac{1}{2}(1+\hat{d}\cdot \bm{\Gamma}),\ \
P^{H}=\frac{1}{2}(1-\hat{d}\cdot \bm{\Gamma})=1-P^{L},\\
&& (P^{L})^{2}=P^{L},\ (P^{H})^{2}=P^{H},\
P^{H}P^{L}=0=P^{L}P^{H}.
\end{eqnarray*}
We can define the covariant gauge field strength, i.e. curvature 
$F_{ab}$ in terms of these
projection operators as
\begin{equation}
F_{ab}=-i\left[ \frac{\partial P^{L}}{\partial d_{a}},\
\frac{\partial P^{L}}{\partial d_{b}} \right] = -i\left[
\frac{\partial P^{H}}{\partial d_{a}},\ \frac{\partial
P^{H}}{\partial d_{b}} \right]\label{Fab}.
\end{equation}
This gauge field is defined over the 5D $\mathbf{d}$ space, with spatial
indices $a,b=1,2,3,4,5$. It is a $4\times 4$ matrix, being a
linear combination of the SO(5) Lie algebra matrices
$\Gamma^{ab}$. It can be explicitly evaluated as
\begin{eqnarray}
F_{ab} &=&\frac{-i}{4} \left[\frac{\partial \hat{d}_{c}}{\partial
d_{a}}\Gamma^{c}, \ \frac{\partial \hat{d}_{d}}{\partial
d_{b}}\Gamma^{d}\right]
\nonumber \\
 &=&\frac{1}{2d^{2}}(\Gamma^{ab}+\hat{d}_{c}\hat{d}_{b}\Gamma^{ca}
-\hat{d}_{c}\hat{d}_{a}\Gamma^{cb}).
\label{Fabform}
\end{eqnarray}
It can also be written as 
\begin{equation}
F_{ab} =\frac{1}{2d^{2}}P_{ab,cd}\Gamma^{cd}
=\frac{1}{2d^{2}}f_{abef}f_{efcd}\Gamma^{cd},
\end{equation}
where $f_{abgh}$ is given in (\ref{Pabcd}).

The gauge potential corresponding to the gauge field strength
$F_{ab}$ is given by $A_{a}=-\frac{1}{2d^{2}}d_{b}\Gamma^{ab}$.
This can be shown by explicit calculations, using the standard
definition
\begin{equation}
F_{ab}=\frac{\partial A_{b}}{\partial d_{a}}- \frac{\partial
A_{a}}{\partial d_{b}}+i[A_{a},A_{b}].
\end{equation}

{}From $F_{ab}$ we can define the dual field strength $G_{ab}$ by
\begin{equation}
G_{ab} =\frac{1}{2} \{F_{ab},
\hat{d}_{c}\Gamma^{c}\}=\frac{1}{2d^{2}}f_{abcd}\Gamma^{cd},
\label{Gab}
\end{equation}
where we used Eqs.~(\ref{GammaId}) and (\ref{Fabform}).

We now define the gauge field strength for each band as
\begin{eqnarray}
&&F_{ab}^{L}=-iP^{L} \left[ \frac{\partial P^{L}}{\partial
d_{a}},\ \frac{\partial P^{L}}{\partial d_{b}}
\right]\label{FabLPL}
\\
&&F_{ab}^{H}=-iP^{H} \left[ \frac{\partial P^{H}}{\partial
d_{a}},\ \frac{\partial P^{H}}{\partial d_{b}} \right]
\label{FabHPH}
\end{eqnarray}
It is easy to see that
\begin{equation}
F_{ab} = F^L_{ab}+F^H_{ab} \ \ ; \ \ G_{ab} = F^L_{ab}-F^H_{ab}.
\end{equation}
Since $F_{ab}$ and $G_{ab}$ are related to each other by a duality
transformation
\begin{equation}
f_{abcd}G_{cd}=F_{ab},\ f_{abcd}F_{cd}=G_{ab},
\end{equation}
$F^L_{ab}$ and $F^H_{ab}$ are self-dual and
anti-self-dual, in the sense that
\begin{equation}
f_{abcd} F^L_{cd} = F^L_{ab} \ \ ; \ \ f_{abcd} F^H_{cd} = -
F^H_{ab}.
\end{equation}
We can explicitly see that $F^L_{ab}$ and $F^{H}_{ab}$ describes 
a gauge field strength with Yang monopole at $\mathbf{d}=0$.
Let us define the two-form $F^{L}$ and $F^{H}$ as
\begin{equation}
F^{L}=\frac{1}{2}F^{L}_{ab}dd_{a}\wedge dd_{b},\  
F^{H}=\frac{1}{2}F^{H}_{ab}dd_{a}\wedge dd_{b}.
\end{equation}
One can calculate that
\begin{eqnarray}
&&\text{tr}(F^{L}\wedge F^{L})=-\text{tr}(F^{H}\wedge F^{H})\nonumber \\
&&\ \ =
\frac{1}{8d^{5}}\epsilon_{abcde}d_{a}\cdot dd_{b}\wedge
dd_{c}\wedge
dd_{d}\wedge
dd_{e}.
\end{eqnarray}
When this is integrated on a four-dimensional hypersurface surrounding 
$\mathbf{d}=0$, it gives the second Chern number multiplied by $8\pi^{2}$.
Therefore $F^{L}$ and $F^{H}$ describe a gauge field with 
the Yang monopole at the origin, with its strength (i.e. the second Chern 
number) given by $+1$ and $-1$, respectively \cite{demler1999}.

Because of the projection operators $P^L$ and $P^H$, $F_{ab}^{L}$
and $F_{ab}^{H}$ can be expressed as SU(2) matrices operating
within the LH and the HH bands respectively. In fact, we can see
that they agree exactly with the conventional definitions of the
non-abelian holonomy or the SU(2) Berry connection. In the
conventional definition, the SU(2) gauge field in the LH band as
$(A_{a}^{L})_{\alpha\beta}=-i\langle \alpha L\mathbf{k}|\frac{\partial
(\beta L\mathbf{k})}{
\partial d_{a}}\rangle$ and its field strength is
$F_{ab}^{L}=\partial_{a}A_{b}^{L}-\partial_{b}A_{a}^{L} +i[A_{a}^{L},\
A_{b}^{L}]$, 
where $\alpha,\beta=1,2$ characterize 
two eigenvectors forming the basis of the LH subspace. 
$A_{a}^{H}$ and $F_{ab}^{H}$ can be defined in a similar
way. The proof of the equivalence between the conventional
definition and the definition (\ref{FabLPL}) can be seen in the
following way, which is essentially the same as in
Ref.~\onlinecite{shankar1994};
\begin{eqnarray}
&&P^{L} \frac{\partial P^{L}}{\partial d_{a}}\frac{\partial
P^{L}}{\partial d_{b}} =- (P^{L} )^{2}\frac{\partial
P^{H}}{\partial d_{a}}\frac{\partial P^{L}}{\partial d_{b}} \nonumber \\
&&\ \ \ =
P^{L} \frac{\partial P^{L}}{\partial d_{a}}P^{H}
\frac{\partial P^{L}}{\partial d_{b}}
= P^{L} \frac{\partial P^{L}}{\partial d_{a}}(P^{H})^{2}
\frac{\partial P^{L}}{\partial d_{b}}
\nonumber \\
&&\ \ \ = -P^{L} \frac{\partial
P^{L}}{\partial d_{a}}P^{H} \frac{\partial P^{H}}{\partial
d_{b}}P^{L}= P^{L} \frac{\partial
P^{L}}{\partial d_{a}}P^{H} \frac{\partial P^{L}}{\partial
d_{b}}P^{L}
\nonumber \\
&&\ \ \ =\sum_{\alpha,\beta} |\alpha
L\rangle\left\langle\frac{\partial (\alpha L)}{\partial
d_{a}}\right| P^{H}\left|\frac{\partial (\beta L)}{\partial
d_{b}}\right\rangle\langle\beta L|,
\end{eqnarray}
where $|\alpha L\rangle=|\alpha L\mathbf{k}\rangle$ and so forth.
Then it follows that
\begin{eqnarray}
&&P^{L}\left( \frac{\partial P^{L}}{\partial d_{a}}\frac{\partial
P^{L}}{\partial d_{b}}
-(a \leftrightarrow b)\right)\nonumber\\
&&\ \ \ =\sum_{\alpha,\beta} |\alpha
L\rangle\left\langle\frac{\partial (\alpha L)}{\partial
d_{a}}\right| P^{H} \left|\frac{\partial (\beta L )}{\partial
d_{b}}\right\rangle\langle\beta L|-(a\leftrightarrow b)
\nonumber \\
&&\ \ \ =\sum_{\alpha,\beta} |\alpha
L\rangle\left\langle\frac{\partial (\alpha L)}{\partial
d_{a}}\right| \left.\frac{\partial (\beta L)}{\partial
d_{b}}\right\rangle\langle\beta
L|\nonumber \\
&&\ \ \ \ \ \ - \sum_{\alpha,\beta,\gamma} |\alpha
L\rangle\left.\left\langle\frac{\partial (\alpha L
)}{\partial
d_{a}}\right| \gamma L
\right\rangle\left\langle \gamma L \left|\frac{\partial
(\beta L)}{\partial d_{b}}\right\rangle\right.\langle\beta L
|\nonumber \\
&&\ \ \ \ \ \ \ \
-(a\leftrightarrow b)\nonumber \\
&&\ \ \ =i\sum_{\alpha,\beta} |\alpha
L\rangle(F_{ab}^{L})_{\alpha\beta}\langle \beta L|,
\end{eqnarray}
which establishes the equivalence between (\ref{FabLPL}) and the
conventional definition of the gauge fields, for example, those
used in Refs.~\onlinecite{demler1999,zhang2001}. The equivalence
between (\ref{FabHPH}) and the conventional definitions can be
shown in a similar way.

{}From these 5D monopole gauge fields, one can easily obtain the 3D
monopole gauge fields by the pull-back mapping. For example,
\begin{equation}
G_{ij}=\frac{\partial d_{a}}{\partial k_{i}} \frac{\partial
d_{b}}{\partial k_{j}} G_{ab} \equiv G_{ij}^{cd} \Gamma^{cd}.
\end{equation}
Substituting the definition of $G_{ab}$ as given in (\ref{Gab}) we
see easily that $G_{ij}^{cd}$ is given by Eq. (\ref{Gij}).

Calculation of $G^{ab}_{ij}$ and $F^{ab}_{ij}$ is straightforward
but somewhat cumbersome. By using Mathematica, we obtain
\begin{eqnarray}
&&F^{ab}_{ij} =\frac{1}{k^{6}}
\epsilon_{ijl}k_{l}(16(\mathbf{k}\cdot\bm{\eta})^{3}+4k^{2}
\mathbf{k}\cdot\bm{\eta})_{ab},
\label{Fabij}\\
&&G^{ab}_{ij} =\frac{1}{2k^{6}}
\epsilon_{ijl}k_{l}(16(\mathbf{k}\cdot\bm{\eta})^{3}+k^{2}
\mathbf{k}\cdot\bm{\eta})_{ab},
\label{Gabij}
\end{eqnarray}
where $\bm{\eta}=(\eta^{x},\eta^{y},\eta^{y})$ and $\eta^{i}$ is
regarded as a 5$\times$5 spin-2 representation of the SU(2) Lie
algebra, satisfying the commutation relation (\ref{eta-algebra}).
In these formulae, $F_{ab}$ and $G_{ab}$ are written in terms of
$\Gamma$ matrices. Alternatively, we can write them in terms of the spin
matrices $S^{i}$;
\begin{eqnarray}
&&F_{ij}=\lambda \left(2\lambda^{2}-\frac{7}{2}\right)
\epsilon_{ijl}\frac{k_{l}}{k^{3}},\label{Fij-lambda}
\\
&&G_{ij}=\lambda \left(\lambda^{2}-\frac{13}{4}\right)
\epsilon_{ijl}\frac{k_{l}}{k^{3}},\label{Gij-lambda}
\end{eqnarray}
where $\lambda=\hat{k}\cdot\mathbf{S}$ is the helicity matrix
\cite{murakami2003}. Eq.~(\ref{Fij-lambda}) has been obtained in
Ref.~\onlinecite{zee1988}; one can show Eq.~(\ref{Gij-lambda}) in
the similar way. Equivalence between (\ref{Fij-lambda}),
(\ref{Gij-lambda}) and (\ref{Fabij}), (\ref{Gabij}) can be shown
by substituting $S^{i}=\eta_{ab}^{i}\Gamma^{ab}$ and using
(\ref{Gammaab-Gammacd}). From (\ref{Fij-lambda}) and
(\ref{Gij-lambda}), we get
\begin{equation}
F_{ij}^{H}=\lambda\left(\frac{\lambda^{2}}{2}-\frac{1}{8}\right)\epsilon_{ijl
}\frac{k_{l}}{k^{3}},\
F_{ij}^{L}=\lambda\left(\frac{3\lambda^{2}}{2}-\frac{27}{8}\right)\epsilon_{ijl
}\frac{k_{l}}{k^{3}}.
\end{equation}
As is expected, $F_{ij}^{H}=0$ for the LH band ($\lambda=\pm 1/2$),
and $F_{ij}^{L}=0$ for the HH band ($\lambda=\pm 3/2$),
This is the field strength of the U(1) (Dirac) monopole with 
monopole strength $\pm 3$ for $\lambda=\pm 3/2$ (HH band) and
$\mp 3$ for $\lambda=\pm 1/2$ (LH band).

Finally we would like to establish the exact equivalence between
the gauge fields introduced above and the Yang-Mills instanton in
Euclidean four-space\cite{belavin1975A} or the Yang monopole gauge
fields over the four-sphere\cite{yang1978A}. The proof essentially
follows that of Jackiw and Rebbi\cite{jackiw1976A}. The 2-form
SO(5) gauge field on $R^{5}$ can be converted to SO(4) 2-form
gauge field on $R^{5}$ by gauge transformation $U$ such that:
\begin{equation}
U^{\dagger}\hat{d}_{a}\Gamma^{a}U=\Gamma^{5}.
\end{equation}
For example, we can take
\begin{equation}
U=\frac{1+\hat{d}_{5}+i\sum_{a=1}^{4}
\Gamma^{a5}\hat{d}_{a}}{\sqrt{2(1+\hat{d}_{5})}}.\end{equation}
By this gauge transformation, the gauge field $A_{a}$ and the
field strength $F_{ab}$ are transformed to
\begin{eqnarray}
&&U^{\dagger}A_{a}U-iU^{\dagger}
\frac{\partial U}{\partial d_{a}} =\tilde{A}_{a},\\
&&U^{\dagger}F_{ab}U=\tilde{F}_{ab}.
\end{eqnarray}
These quantities $\tilde{A}_{a}$ and $\tilde{F}_{ab}$ are linear
combinations of $\Gamma^{mn}$ $(m,n=1,2,3,4)$, belonging to the
SO(4) algebra. Explicitly they are written as
\[
\tilde{A}_{a}=-\frac{1}{2d(1+\hat{d}_{5})}
\sum_{b=1}^{4}\hat{d}_{b}\Gamma^{ab}\ \ (a=1,2,3,4),
\ \ \tilde{A}_{5}=0, 
\]
\[
\tilde{F}_{a5}=-\frac{1}{2d^{2}}\sum_{b=1}^{4}\hat{d}_{b}\Gamma^{ab}\ \
(a=1,2,3,4), 
\]
\begin{eqnarray*}
&&
\tilde{F}_{ab}=\frac{1}{2d^{2}}\left(
\Gamma^{ab}-\frac{1}{1+\hat{d}_{5}}\sum_{c=1}^{4}
\hat{d}_{c}(\hat{d}_{b}\Gamma^{ac}-\hat{d}_{a}\Gamma^{bc})\right)\\
&& \makebox[3cm]{}(a,b=1,2,3,4),
\end{eqnarray*}
which are exactly the SO(4)=SU(2)$\times$SU(2) gauge fields used in
the context of 4DQHE\cite{zhang2001}.

\begin{acknowledgments}
We thank A. Bernevig, C. H. Chern, D. Culcer, J. P. Hu, 
T. Jungwirth, A. H. MacDonald, 
Q. Niu, N. A. Sinitsyn, J. Sinova  and C. J. Wu for helpful discussions and for
sharing results prior to their publications. This work is
supported by Grant-in-Aids from the Ministry of Education,
Culture, Sports, Science and Technology of Japan, the US NSF under
grant numbers DMR-9814289, and the US Department of Energy, Office
of Basic Energy Sciences under contract DE-AC03-76SF00515.
\end{acknowledgments}


\end{document}